\documentstyle[12pt]{article}
\begin{document}
\def\rr#1{$^{#1}$}
\newcommand{\sect}[1]{\setcounter{equation}{0}\section{#1}}
%
%
%
%
\def\ggg{{\eufm g}}
\def\hb{{\bfmath h}}
\def\ie{{\it i.e.\/}}
\def\eg{{\it e.g.\/}}
\def\gg{{\>\widehat{g}\>}}
\def\GG{{\>\hat{G}\>}}
\def\ss{{\>\widehat{s}\>}}
\def\sb{{\bf s}}
\def\sw{{\bf s}^w}
\def\Heis{{\cal H}[w]}
\def\ad{{\rm ad\>}}
\def\Ker{{\rm Ker\/}}
\def\Im{{\rm Im\/}}
\def\Pos{{\rm P}_{\geq0[\sw]}}
\def\Neg{{\rm P}_{<0[\sw]}}
\def\W{$\cal W$}
\def\cl{{\cal L}}
\def\pa{\partial}
\mathchardef\bphi="731E
\mathchardef\balpha="710B
\mathchardef\bomega="7121
\def\alb{{\bfmath\balpha}}
\def\pp{{\bfmath\bphi}}
\def\omb{{\bfmath\bomega}}
\def\Hb{{\bfmath H}}

\def\sect#1{\section{#1}}

\def\rf#1{(\ref{#1})}
\def\lab#1{\label{#1}}
\def\nonu{\nonumber}
\def\br{\begin{eqnarray}}
\def\er{\end{eqnarray}}
\def\be{\begin{equation}}
\def\ee{\end{equation}}
\def\eq{\!\!\!\! &=& \!\!\!\! }
\def\foot#1{\footnotemark\footnotetext{#1}}
\def\lb{\lbrack}
\def\rb{\rbrack}
\def\llangle{\left\langle}
\def\rrangle{\right\rangle}
\def\blangle{\Bigl\langle}
\def\brangle{\Bigr\rangle}
\def\llbrack{\left\lbrack}
\def\rrbrack{\right\rbrack}
\def\lcurl{\left\{}
\def\rcurl{\right\}}
\def\({\left(}
\def\){\right)}
\newcommand{\nit}{\noindent}
\newcommand{\ct}[1]{\cite{#1}}
\newcommand{\bi}[1]{\bibitem{#1}}
\def\lskip{\vskip\baselineskip\vskip-\parskip\noindent}
\relax

\def\tr{\mathop{\rm tr}}
\def\Tr{\mathop{\rm Tr}}
\def\v{\vert}
\def\bv{\bigm\vert}
\def\Bgv{\;\Bigg\vert}
\def\bgv{\bigg\vert}
\newcommand\partder[2]{{{\partial {#1}}\over{\partial {#2}}}}
\newcommand\funcder[2]{{{\delta {#1}}\over{\delta {#2}}}}
\newcommand\Bil[2]{\Bigl\langle {#1} \Bigg\vert {#2} \Bigr\rangle}  
\newcommand\bil[2]{\left\langle {#1} \bigg\vert {#2} \right\rangle} 
\newcommand\me[2]{\left\langle {#1}\bv {#2} \right\rangle} 
\newcommand\sbr[2]{\left\lbrack\,{#1}\, ,\,{#2}\,\right\rbrack}
\newcommand\pbr[2]{\{\,{#1}\, ,\,{#2}\,\}}
\newcommand\pbbr[2]{\lcurl\,{#1}\, ,\,{#2}\,\rcurl}
%
\def\a{\alpha}
\def\at{{\tilde A}^R}
\def\atc{{\tilde {\cal A}}^R}
\def\atcm#1{{\tilde {\cal A}}^{(R,#1)}}
\def\b{\beta}
\def\btil{{\tilde b}}
\def\bra#1{\langle #1 \mid}
\def\ket#1{\mid #1 \rangle}
\def\dc{{\cal D}}
\def\d{\delta}
\def\D{\Delta}
\def\eps{\epsilon}
\def\vareps{\varepsilon}
\def\fptil{{\tilde F}^{+}}
\def\fmtil{{\tilde F}^{-}}
\def\gh{{\hat g}}
\def\g{\gamma}
\def\G{\Gamma}
\def\grad{\nabla}
\def\h{{1\over 2}}
\def\l{\lambda}
\def\L{\Lambda}
\def\m{\mu}
\def\n{\nu}
\def\o{\over}
\def\om{\omega}
\def\O{\Omega}
\def\p{\phi}
\def\P{\Phi}
\def\pa{\partial}
\def\pr{\prime}
\def\pt{{\tilde \Phi}}
\def\qs{Q_{\bf s}}
\def\ra{\rightarrow}
\def\s{\sigma}
\def\S{\Sigma}
\def\t{\tau}
\def\th{\theta}
\def\Th{\Theta}
\def\tpp{\Theta_{+}}
\def\tmm{\Theta_{-}}
\def\tpg{\Theta_{+}^{>}}
\def\tms{\Theta_{-}^{<}}
\def\tp0{\Theta_{+}^{(0)}}
\def\tm0{\Theta_{-}^{(0)}}
\def\ti{\tilde}
\def\wti{\widetilde}
\def\jc{J^C}
\def\bj{{\bar J}}
\def\sj{{\jmath}}
\def\bsj{{\bar \jmath}}
\def\bp{{\bar \p}}
\def\vp{\varphi}
\def\vt{{\tilde \varphi}}
\def\faa{Fa\'a di Bruno~}
\def\ca{{\cal A}}
\def\cb{{\cal B}}
\def\ce{{\cal E}}
\def\cg{{\cal G}}
\def\cgh{{\hat {\cal G}}}
\def\cgt{{\tilde{\cg}}}
\def\ch{{\cal H}}
\def\chh{{\hat {\cal H}}}
\def\cl{{\cal L}}
\def\cm{{\cal M}}
\def\cn{{\cal N}}
\def\ns{N_{{\bf s}}}
\newcommand\sumi[1]{\sum_{#1}^{\infty}}   
\newcommand\fourmat[4]{\left(\begin{array}{cc}  
{#1} & {#2} \\ {#3} & {#4} \end{array} \right)}

%
\def\lie{{\cal G}}
\def\kmlie{{\hat{\cal G}}}
\def\dlie{{\cal G}^{\ast}}
\def\elie{{\widetilde \lie}}
\def\edlie{{\elie}^{\ast}}
\def\hlie{{\cal H}}
\def\flie{{\cal F}}
\def\wlie{{\widetilde \lie}}
\def\f#1#2#3 {f^{#1#2}_{#3}}
\def\winf{{\sf w_\infty}}
\def\win1{{\sf w_{1+\infty}}}
\def\hwinf{{\sf {\hat w}_{\infty}}}
\def\Winf{{\sf W_\infty}}
\def\Win1{{\sf W_{1+\infty}}}
\def\hWinf{{\sf {\hat W}_{\infty}}}
\def\Rm#1#2{r(\vec{#1},\vec{#2})}          
\def\OR#1{{\cal O}(R_{#1})}           
\def\ORti{{\cal O}({\widetilde R})}           
\def\AdR#1{Ad_{R_{#1}}}              
\def\dAdR#1{Ad_{R_{#1}^{\ast}}}      
\def\adR#1{ad_{R_{#1}^{\ast}}}       
\def\KP{${\rm \, KP\,}$}                 
\def\KPl{${\rm \,KP}_{\ell}\,$}         
\def\KPo{${\rm \,KP}_{\ell = 0}\,$}         
\def\mKPa{${\rm \,KP}_{\ell = 1}\,$}    
\def\mKPb{${\rm \,KP}_{\ell = 2}\,$}    
%
\def\rlx{\relax\leavevmode}
\def\inbar{\vrule height1.5ex width.4pt depth0pt}
\def\IZ{\rlx\hbox{\sf Z\kern-.4em Z}}
\def\IR{\rlx\hbox{\rm I\kern-.18em R}}
\def\IC{\rlx\hbox{\,$\inbar\kern-.3em{\rm C}$}}
\def\IN{\rlx\hbox{\rm I\kern-.18em N}}
\def\IO{\rlx\hbox{\,$\inbar\kern-.3em{\rm O}$}}
\def\IP{\rlx\hbox{\rm I\kern-.18em P}}
\def\IQ{\rlx\hbox{\,$\inbar\kern-.3em{\rm Q}$}}
\def\IF{\rlx\hbox{\rm I\kern-.18em F}}
\def\IG{\rlx\hbox{\,$\inbar\kern-.3em{\rm G}$}}
\def\IH{\rlx\hbox{\rm I\kern-.18em H}}
\def\II{\rlx\hbox{\rm I\kern-.18em I}}
\def\IK{\rlx\hbox{\rm I\kern-.18em K}}
\def\IL{\rlx\hbox{\rm I\kern-.18em L}}
\def\one{\hbox{{1}\kern-.25em\hbox{l}}}
\def\0#1{\relax\ifmmode\mathaccent"7017{#1}%
B        \else\accent23#1\relax\fi}
\def\omz{\0 \omega}
%
\def\ltimes{\mathrel{\vrule height1ex}\joinrel\mathrel\times}
\def\rtimes{\mathrel\times\joinrel\mathrel{\vrule height1ex}}
%
\def\mark{\noindent{\bf Remark.}\quad}
\def\prop{\noindent{\bf Proposition.}\quad}
\def\theor{\noindent{\bf Theorem.}\quad}
\def\name{\noindent{\bf Definition.}\quad}
\def\exam{\noindent{\bf Example.}\quad}
\def\proof{\noindent{\bf Proof.}\quad}
%
%
\def\PRL#1#2#3{{\sl Phys. Rev. Lett.} {\bf#1} (#2) #3}
\def\NPB#1#2#3{{\sl Nucl. Phys.} {\bf B#1} (#2) #3}
\def\NPBFS#1#2#3#4{{\sl Nucl. Phys.} {\bf B#2} [FS#1] (#3) #4}
\def\CMP#1#2#3{{\sl Commun. Math. Phys.} {\bf #1} (#2) #3}
\def\PRD#1#2#3{{\sl Phys. Rev.} {\bf D#1} (#2) #3}
\def\PRv#1#2#3{{\sl Phys. Rev.} {\bf #1} (#2) #3}
\def\PLA#1#2#3{{\sl Phys. Lett.} {\bf #1A} (#2) #3}
\def\PLB#1#2#3{{\sl Phys. Lett.} {\bf #1B} (#2) #3}
\def\JMP#1#2#3{{\sl J. Math. Phys.} {\bf #1} (#2) #3}
\def\PTP#1#2#3{{\sl Prog. Theor. Phys.} {\bf #1} (#2) #3}
\def\SPTP#1#2#3{{\sl Suppl. Prog. Theor. Phys.} {\bf #1} (#2) #3}
\def\AoP#1#2#3{{\sl Ann. of Phys.} {\bf #1} (#2) #3}
\def\PNAS#1#2#3{{\sl Proc. Natl. Acad. Sci. USA} {\bf #1} (#2) #3}
\def\RMP#1#2#3{{\sl Rev. Mod. Phys.} {\bf #1} (#2) #3}
\def\PR#1#2#3{{\sl Phys. Reports} {\bf #1} (#2) #3}
\def\AoM#1#2#3{{\sl Ann. of Math.} {\bf #1} (#2) #3}
\def\UMN#1#2#3{{\sl Usp. Mat. Nauk} {\bf #1} (#2) #3}
\def\FAP#1#2#3{{\sl Funkt. Anal. Prilozheniya} {\bf #1} (#2) #3}
\def\FAaIA#1#2#3{{\sl Functional Analysis and Its Application} {\bf #1} (#2)
#3}
\def\BAMS#1#2#3{{\sl Bull. Am. Math. Soc.} {\bf #1} (#2) #3}
\def\TAMS#1#2#3{{\sl Trans. Am. Math. Soc.} {\bf #1} (#2) #3}
\def\InvM#1#2#3{{\sl Invent. Math.} {\bf #1} (#2) #3}
\def\LMP#1#2#3{{\sl Letters in Math. Phys.} {\bf #1} (#2) #3}
\def\IJMPA#1#2#3{{\sl Int. J. Mod. Phys.} {\bf A#1} (#2) #3}
\def\AdM#1#2#3{{\sl Advances in Math.} {\bf #1} (#2) #3}
\def\RMaP#1#2#3{{\sl Reports on Math. Phys.} {\bf #1} (#2) #3}
\def\IJM#1#2#3{{\sl Ill. J. Math.} {\bf #1} (#2) #3}
\def\APP#1#2#3{{\sl Acta Phys. Polon.} {\bf #1} (#2) #3}
\def\TMP#1#2#3{{\sl Theor. Mat. Phys.} {\bf #1} (#2) #3}
\def\JPA#1#2#3{{\sl J. Physics} {\bf A#1} (#2) #3}
\def\JSM#1#2#3{{\sl J. Soviet Math.} {\bf #1} (#2) #3}
\def\MPLA#1#2#3{{\sl Mod. Phys. Lett.} {\bf A#1} (#2) #3}
\def\JETP#1#2#3{{\sl Sov. Phys. JETP} {\bf #1} (#2) #3}
\def\CAG#1#2#3{{\sl  Commun. Anal\&Geometry} {\bf #1} (#2) #3}
\def\JETPL#1#2#3{{\sl  Sov. Phys. JETP Lett.} {\bf #1} (#2) #3}
\def\PHSA#1#2#3{{\sl Physica} {\bf A#1} (#2) #3}
\def\PHSD#1#2#3{{\sl Physica} {\bf D#1} (#2) #3}
\def\PJA#1#2#3{{\sl Proc. Japan. Acad.} {\bf #1A} (#2) #3}
\def\JPSJ#1#2#3{{\sl J. Phys. Soc. Japan} {\bf #1} (#2) #3}
\def\SJPN#1#2#3{{\sl Sov. J. Part. Nucl.} {\bf #1} (#2) #3}

\begin{titlepage}
\vspace*{-2 cm}
\noindent
December, 1996 \hfill{US-FT/49-96}\\
hep-th/9701006 \hfill{IFT-P.005/97} 

\vskip 1 cm

\begin{center}
{\Large\bf ASPECTS OF SOLITONS IN AFFINE INTEGRABLE HIERARCHIES\footnote{Two 
talks presented by the
authors at the ``International Workshop on Selected Topics of Theoretical and 
Modern Mathematical Physics - SIMI/96'', Tbilisi, Georgia, September/96.}}
\vglue 1  true cm
{\bf Luiz A. FERREIRA}$^1$ and {\bf Joaqu\'\i n S\'ANCHEZ GUILL\'EN}$^2$\\

\vspace{1 cm}

$^1${\footnotesize Instituto de F\'\i sica Te\'orica - IFT/UNESP\\
Rua Pamplona 145\\
01405-900, S\~ao Paulo - SP, Brazil}\\

\vspace{1 cm} 

$^2${\footnotesize Facultad de F\'\i sica,\\
 Universidad de Santiago de Compostela,\\
15706 Santiago de Compostela, Spain.}
\medskip
\end{center}

\normalsize
\vskip 0.2cm

\begin{center}
{\large {\bf ABSTRACT}}\\
\end{center}


\noindent
{\footnotesize 
We consider a very large class of hierarchies of zero-curvature equations
constructed from affine Kac-Moody algebras $\cgh$. 
We argue that one of the basic ingredients for the appearance of soliton 
solutions in such theories is the existence of ``vacuum solutions'' 
corresponding to Lax operators lying in some abelian (up to central term)
subalgebra of $\cgh$. Using the dressing transformation procedure we construct
the solutions in the orbit of those vacuum solutions, and conjecture that the
soliton solutions  correspond to some  special points in
those orbits. The generalized tau-function for those hierarchies are defined
for  integrable highest weight representations of $\cgh$, and it applies for
any level of the representation and it is independent of its realization.  
We illustrate our methods with the recently proposed non abelian Toda models
coupled to matter fields. A very special 
class of such theories possess a
$U(1)$ Noether charge that, under a suitable gauge fixing of the conformal
symmetry, is proportional to a topological charge. That leads to a mechanism
that confines the matter fields inside the solitons.} 
\vglue 1 true cm

\end{titlepage}

\sect{Introduction}

The study of soliton solutions of non linear differential  
equations has been developed considerably in the last decades using 
(apparently) quite diverse methods. In spite of the great variety of types of
equations considered, some basic features seem to be common to a large class of
them. Attempts to unify such various aspects of basic classical Soliton Theory 
is clearly very important and may lead to new insights into the role of
solitons in Physics and Mathematics. 

Among the several methods of constructing solutions for non linear differential
equations we have the Hirota method \ct{HIR}, the dressing transformation
procedure \ct{dress1,dress2,dress3,BABELON,BABT}, 
Backlund transformations \ct{bac,liao}, the inverse scattering method
\ct{fad}, the Leznov-Saveliev algebraic construction \ct{LS92} and the
tau-function approach \ct{KW,dress2}. Each one of these methods have their own
advantages, and the choice of  one or the other depends on the 
particular problem and model one wants to address. However, soliton
solutions are special and the theories presenting them must possess some common
structures. By solitons here we mean a solution localized in space that travels
without dispersion, and that keeps its form when scattered by other soliton,
suffering just a shift in its position with respect to the one it would have if
not for the scattering. In fact, in some theories like the Toda models which
possess two dimensional Lorentz invariance, the solitons can have a particle
interpretation and there are indications for the existence of a duality between
the one-soliton solutions and the fundamental particles. In four dimensions,
a similar duality, generalizing the electromagnetic duality, seems to
exist between monopoles and gauge particles \ct{duality}. 

Pratically all theories in one and two space time dimensions, presenting 
soliton solutions, have a representation in terms of a zero curvature 
condition or Lax-Zakharov-Shabat equation. In addition, the corresponding
Lax operators lie in some infinite dimensional Lie algebra. In fact, we can say
that basically almost all known soliton equations are related to 
Kac-Moody algebras \ct{DS,GEN}. 

In this paper we try to unify some aspects concerning soliton solutions. We
argue that a basic ingredient for the appearence of soliton solutions is that
there must exist one or several solutions, which we call ``vacuum solutions'',
such that the Lax operators, when evaluated on them, should lie in some abelian
subalgebra (up to central term) of the  Kac-Moody algebra associated to the
model. Such subalgebra can be written as an algebra of oscillators $b_i$  
\be
[b_i,b_j] = i \>\beta_i\>C\> \delta_{i+j,0}
\lab{oscila}
\ee
with $\beta_i$ being some complex numbers and $C$ the central element of the
Kac-Moody algebra $\cgh$. In several cases, but not in all, \rf{oscila} 
constitutes a Heisenberg subalgebra \ct{kacpet} of $\cgh$. In addition, we 
argue 
that 
the components of the
Lax operators in the direction of the $b_i$'s should be constant for the vacuum
solutions. 

Using the dressing transformation method one can then construct, out of a given
vacuum solution, an orbit of
solutions  parametrized by a constant group element $\rho$ of the Kac-Moody
group. We conjecture that the soliton solutions of those theories lie on such
orbits, and that they correspond to points of the orbits where $\rho$ is the
product of exponentials of eigenvectors of the constant elements defined by the
Lax operators evaluated on the corresponding vacuum solution. Such observations
provide not only a powerful and elegant method of constructing soliton
solutions, but also allow us to connect and generalize several results known in
the literature \ct{WILa,WILb}. The so called solitonic specialization  of the
Leznov-Saveliev solution proposed in \ct{TUROLA,TUROLB,SOLSPEC} in the context 
of Toda type models,
 can then be connected to the
dressing method applied to such vacuum solutions. In addition, we believe that
in several cases there is a connection with the Backlund transformation method.

We can also connect that observation with tau-functions. 
We define 
tau-functions as states in integrable highest weight representations of
the Kac-Moody algebra, lying in the orbit, under the
action of the group elements performing the dressing transformations,  
of the highest weight state. Our
definition is independent of the level of the representation and also on the
way it is realised, constituting a generalization of the
previous definitions of tau-functions for level one vetex operator 
representations \ct{KW}. The connection with the Hirota method is then made by
realizing the Hirota's tau-function as projections of the tau-function on some 
suitable states of representation. The truncation of the Hirota's expansion is
then understood in terms of the nilpotency of some operators in those 
integrable representations. 

We illustrate our methods with the recently proposed non abelian Toda models
coupled to matter fields \ct{SAVGERV}. In fact, we discuss a very special class 
of 
such theories where the solitons play a crucial role. These models possess a
$U(1)$ Noether charge that, under a suitable gauge fixing of the conformal
symmetry, is proportional to a topological charge. That leads to a mechanism
that confines the matter fields inside the solitons. 

The paper contains the material presented by the authors in their talks at the 
``International Workshop on Selected Topics of Theoretical and Modern
Mathematical Physics - SIMI/96'', held in Tbilisi, Georgia (September/96), and 
summarizes results of references \ct{tau,SAVGERV,fms}. It 
is organized as follows. In section 2 we introduce the type of hierarchies of
soliton equations we shall consider, and discuss their vacuum solutions. In
section 3 we define the dressing transformations and in section 4 we construct
the soliton solutions. Section 5 introduces the tau-functions and discuss how
they connect to previous definitions and to the Hirota's tau functions. The non
abelian Toda models are introduced in section 6, their soliton solutions
constructed in section 7, and their properties discussed in sections 8 and 9.
In section 10 we present in great detail, the example of the model associated 
to the principal gradation of $sl(2)^{(1)}$.

\sect{Hierarchies and vacuum solutions}

Non-linear integrable hierarchies of equations are most conveniently
discussed by associating them with a system of first-order
differential equations
\be
\cl_N \Psi = 0\>,
\lab{LinProb}
\ee
where $\cl_N $ are Lax operators of the form
\be
\cl_N  \equiv {\pa \, \over \pa t_N} - A_N 
\lab{LaxGen}
\ee
and the variables $t_N$ are the various ``times'' of the hierarchy. Then, the
equivalent zero-curvature formulation is obtained through the integrability 
conditions of the associated linear problem~\rf{LinProb},
\be
[{\cal L}_N\>, \>{\cal L}_M]=0\>. 
\lab{ZeroCurv}
\ee
An equivalent way to express the relation between the solutions of the
zero-curvature equations and of the associated linear problem is
\be
A_N = {\partial \Psi\over \partial t_{N}}\>
\Psi^{-1}\>.
\lab{psidef}
\ee

The class of integrable hierarchies of zero-curvature equations that will be
studied here is constructed from graded Kac-Moody
algebras in the following way. Consider a complex affine
Kac-Moody algebra 
${\tilde \cg}  = \cgh + \IC D$ , associated to a simple finite Lie
algebra $\cg$ of rank $r$, and an integer gradation of its
derived algebra $\cgh$ labelled by a vector $\sb=(s_0,s_1,\ldots ,s_r)$ of
$r+1$ non-negative co-prime integers such that 
\be
\cgh= \bigoplus_{i\in{\IZ}} \cgh_i(\sb)\> \quad {\rm
and}\quad [\cgh_i(\sb),\cgh_j(\sb)] \subseteq \cgh_{i+j}(\sb)\>. 
\lab{Gradation}
\ee

According to \ct{kac1}, integral gradations of $\cgh$ are labelled by a
set of co-prime integers  ${\bf s}=\( s_0,  s_1, \ldots s_r \)$, and
the grading operators are given by
\be
Q_{{\bf s}} \equiv H_{{\bf s}} + N_{{\bf s}}\, D - {1\o {2 N_{{\bf s}}}} 
\Tr \( H_{{\bf s}} \)^2 \, C \, .
\lab{gradop}
\ee
Here
\be
H_{{\bf s}} \equiv \sum_{a=1}^{r} s_a \l^v_a \cdot H^0 \, , \qquad
N_{{\bf s}} \equiv \sum_{i=0}^{r} s_i m_i^{\psi} \, , \qquad 
\psi = \sum_{a=1}^{r}  m_a^{\psi} \a_a \, , \quad m_0^{\psi} = 1\, ;
\ee
$H^0$ is an element of the Cartan subalgebra of $\cg$; $\a_a$,
$a=1,2,\ldots r$, are its simple roots; $\psi$ is its maximal root; 
$m_a^{\psi}$ the integers in  expansion $\psi = \sum_{a=1}^r m_a^{\psi} \a_a$; 
and $\l^v_a
$ are the fundamental co--weights satisfying the relation $\a_a \cdot \l^v_b =
\d_{ab}$.

We have in mind basically two types of integrable systems. The first one
corresponds to the Generalized Drinfel'd-Sokolov Hierarchies considered
in~\ct{GEN}, and~\ct{TAUTIM}, which are generalizations of the KdV type
hierarchies studied in~\ct{DS}. In particular, and using the parlance of
the original references, we will be interested in the generalized mKdV
hierarchies, whose construction can be summarised as follows
(see~\ct{GEN} and, especially,~\ct{TAUTIM} for details). Given an
integer gradation
$\sb$ of
$\cgh$ and a semisimple constant element $E_l$ of grade $l$ with respect to
$\sb$, one defines the Lax  operator 
\be
L \equiv \pa_x + E_l + A\>,
\lab{xlax}
\ee
where the components of $A$ are the fields of the
hierarchy. In~\ct{TAUTIM}, it was shown that the component of $A$ along
the central term of~$\cgh$ should not be considered as an actual degree of
freedom of the hierarchy. This is the reason why these hierarchies can be
equivalently formulated both in terms of affine Kac-Moody algebras or of the
corresponding loop algebras. 
 They are functions of
$x$ and of the other times of the hierarchy taking values in the subspaces  of
$\cgh$ with grades ranging from $0$ to $l-1$. For each element in the centre of
$\Ker(\ad E_l )$ with positive $\sb$-grade $N$, one constructs a local
functional of those fields,
$B_N$, whose components take values in the subspaces
$\cgh_0(\sb), \ldots, \cgh_N(\sb)$. Then, $B_N$ defines the flow equation
\be
{\pa L\over \pa t_N}\> =\>  \bigl[ B_N \> , \> L \bigr]\>,
\lab{flow}
\ee
and the resulting Lax operators $\cl_N = {\pa /\pa t_N}  - B_N$ commute among 
themselves~\ct{GEN}.

The second type of integrable systems corresponds to the non-abelian affine
Toda theories~\ct{LS,fms,SAVGERV,UNDER}, and a very general class of these
models will be described in section \ref{sec:formulation}. 

An important common feature of all those hierarchies is that they possess
trivial solutions which will be called ``vacuum solutions''. These 
particular solutions are singled out by the condition that the  Lax operators
evaluated on them lie on some abelian subalgebra of~$\cgh$ , up to central 
terms.
Then, the dressing  transformation method can be used to generate an orbit of
solutions out of each ``vacuum''. Moreover, it is generally conjectured that
multi-soliton solutions lie  in the resulting orbits. As a bonus, the fact that
we only consider the particular subset of solutions connected with a generic
vacuum  allows one to perform the calculations in a  very general way and,
consequently, our results  apply to a much broader class of hierarchies. 

For a given choice of the Kac-Moody algebra $\cgt$ and the gradation $\sb$, let 
us
consider Lax  operators of the form \rf{LaxGen} where the potentials can be
decomposed as
\be
A_N = \sum_{i=N_{-}}^{N_{+}} A_{N,i} \> , \quad  {\rm where}\quad
A_{N,i}  \in \cgh_i(\sb)
\lab{genpot}
\ee
$N_{-}$ and $N_{+}$ are non-positive and non-negative integers, 
respectively, and the times $t_N$ are labelled by (positive or
negative) integer numbers. The particular form of these potentials will be
constrained only by the condition that the corresponding hierarchy admits 
vacuum
solutions where they take the form  
\be
A_N^{({\rm vac})}\> =\>  \sum_{i=N_{-}}^{N_{+}} c_N^i b_i+ f_N (t)\> c
\>\equiv  \> \varepsilon_N + f_N (t)\> C\>. 
\lab{vacpot}
\ee
In this equation, $C$ is the central element of $\cgh$, and  
$b_i\in \cgh_i(\sb)$ 
are the generators of a subalgebra $\ss $ of $\cgh$ defined by
\be
\ss = \{b_i \in\cgh_i(\sb)\>,\; i\in E\subset {\IZ}\bigm| [b_i,b_j] =
i \>\beta_i\>C\> \delta_{i+j,0}\}\>,
\lab{Commute}
\ee
where  $\beta_i$ are arbitrary (vanishing or non-vanishing) complex numbers
such that $\beta_{-i} = \beta_i$, and
$E$ is some set of integers numbers. Moreover, $c_N^i$ are also arbitrary
numbers, and $f_N (t)$ are $\IC$-functions of the times $t_N$ that satisfy
the equations
\be
{\pa\> f_N(t)\over \pa\> t_M} \> -\> {\pa\> f_M(t)\over \pa\> t_N}\> = \>
\sum_{i}\> i\> \beta_i\> c_M^i\> c_N^{-i}\>. 
\lab{ZeroTriv}
\ee 

These vacuum potentials correspond to the solution of the associated linear
problem given by the group  element~\rf{psidef} 
\be
\Psi^{({\rm vac})} = \exp \( \sum_{N} \varepsilon_N t_N
\> +\> \gamma (t) \> C \)
\lab{vacelem}
\ee
where the numeric function $\gamma (t)$ is a solution of the equations
\be
{\pa \gamma (t) \over \pa t_N} = f_N (t) + {1\over 2}\sum_{M,i}
i\>\beta_i\> c_N^i\> c_M^{-i}\>t_M\>. 
\lab{gamma}
\ee

\sect{Dressing Transformations}
\label{sec:dressing}

In terms of the associated linear problem, one can define an important set of
transformations called ``dressing transformations'', which take known solutions
of the hierarchy to new solutions. Regarding the structure of the integrable
hierarchies, these transformations have a deep meaning and, in fact,
the group of dressing transformations can be viewed as the classical precursor
of the quantum group symmetries~\ct{BABELON}. Denote by $\GG_-(\sb)$, 
 $\GG_+(\sb)$, and $\GG_0(\sb)$ the subgroups of the Kac-Moody group $\GG$ 
 formed
by exponentiating the subalgebras $\cgh_{<0}(\sb ) \equiv \bigoplus_{i<0}
\cgh_i(\sb)$, $\cgh_{>0}(\sb)\equiv \bigoplus_{i>0} \cgh_i(\sb)$, and 
$\cgh_0(\sb)$,
respectively. According to Wilson~\ct{WILa,WILb}, the dressing
transformations can be described in the following way. Consider a solution
$\Psi$ of the linear problem~\rf{LinProb}, and let 
$\rho= \rho_-\> \rho_0\> \rho_+$ be a constant element in the ``big cell'' of  
$\GG$, {\it
i.e.\/}, in the subset $\GG_-(\sb)\> \GG_0(\sb)\> \GG_+(\sb)$ of $\GG$, such   
that 
\be
\Psi\> \rho\> \Psi^{-1} = (\Psi\> \rho\> \Psi^{-1})_{<0}\> (\Psi\> \rho\>
\Psi^{-1})_0\> (\Psi\> \rho\> \Psi^{-1})_{>0}\>.
\lab{Factor}
\ee
Notice that these conditions are equivalent to say that both $\rho$ and $\Psi\> 
\rho\> \Psi^{-1}$ admit a generalized Gauss decomposition with respect to the
gradation $\sb\>$. Define 
\br
\Psi^{\rho} \> & = &\> {\tm0} [(\Psi\> \rho \> \Psi^{-1})_{<0}]^{-1}\> \Psi 
\rho\nonu\\
&  = &\> {\tp0} \> (\Psi\> \rho\> \Psi^{-1})_{>0} \>\Psi\> 
\lab{Dressing}
\er
where
\be
{{\tm0}}^{-1}\, {\tp0}  = \( \Psi \rho \Psi^{-1}\)_{0}.
\lab{tp0tm0}
\ee

Then, $\Psi^{\rho}$ is another solution of the linear problem. In order to 
prove it, introduce the
notation $g_{\pm} \equiv (\Psi\> \rho\> \Psi^{-1})_{\pm}$ and $\partial_N
\equiv \partial/\partial t_N$, and consider
\br
\partial_N \Psi^{\rho}\> {\Psi^{\rho}}^{-1} & = & 
\partial_N {\tm0} \> {\tm0}^{-1}
- {\tm0} g_-^{-1}\> \partial_N g_- {\tm0}^{-1}  \nonu\\
&+&  
{\tm0} g_-^{-1} (\partial_N \Psi \> \Psi^{-1}) g_- {\tm0}^{-1}\nonu\\
& = &\partial_N {\tp0} \> {\tp0}^{-1} + 
{\tp0} \partial_N g_+ \> g_+^{-1} {\tp0}^{-1} \nonu\\
&+& 
{\tp0} g_+ (\partial_N \Psi\> \Psi^{-1}) g_+^{-1} {\tp0}^{-1}
\er
Then, the first identity implies that $\partial_N \Psi^{\rho}\> 
{\Psi^{\rho}}^{-1} \in 
\bigoplus_{i\leq N_+}\cgh_{i}(\sb)$, and the second that $\partial_N 
\Psi^{\rho}\> {\Psi^{\rho}}^{-1} 
\in \bigoplus_{i\geq N_-}\cgh_{i}(\sb)$. Consequently
\be
A_{N}^{\rho} = {\partial \Psi^{\rho}\over \partial t_N}\> ({\Psi^{\rho}})^{-1}
\in \bigoplus_{i=N_-}^{N_+}\cgh_{i}(\sb)\>,
\lab{DressZero}
\ee
and, taking into account~\rf{genpot}, it is a solution of the hierarchy of
zero-curvature equations. If the fields of the hierarchy are such that
$A_{N,i}$ does not span the whole subspace $\cgh_{i}(\sb)$ then we have to
impose further constraints on the group elements performing the dressing
transformation. 

For any $\rho$ lying in the big cell of $\GG$, the transformation 
\be
{\cal D}_{\rho} : \Psi\> \mapsto\> \Psi^{\rho}\>, \quad{\rm or}\quad A_{N} 
\mapsto A_{N}^{\rho}\>,
\lab{DresTr}
\ee 
is called a dressing transformation.

\sect{Solitons out of vacuum solutions}
\label{sec:soliton}

We now consider the orbit of the vacuum solution~\rf{vacelem} under the group  
of dressing transformations. 
For any element $\rho$ of the big cell of $\GG$, let us define
\be
\tpg = \( \Psi^{({\rm vac})} \rho {\Psi^{({\rm vac})}}^{-1}\)_{>0} \, , \qquad 
\tms = \( \Psi^{({\rm vac})} \rho {\Psi^{({\rm vac})}}^{-1}\)_{<0}^{-1}.
\lab{tpms2}
\ee
and 
\be
\tpp = {\tp0} \,\tpg \, , \qquad \qquad \tmm = {\tm0} \,\tms ;
\lab{tpm}
\ee   
where $\tp0$ and $\tm0$ are the same as in \rf{tp0tm0}, but with $\Psi$
replaced by $\Psi^{({\rm vac})}$ defined in \rf{vacelem}. 

Now, the orbit of the vacuum solution~\rf{vacelem} can be easily constructed 
using Eqs.~\rf{Dressing} and
\rf{DressZero}. Under the dressing transformation generated by $\rho$,
\be
\Psi^{(\rm vac)}\> \mapsto\> \Psi^{\rho} = {\tmm}\> 
\Psi^{(\rm vac)} \rho = 
{\tpp}\> \Psi^{(\rm vac)}\> \>,
\lab{Simple}
\ee
or, equivalently, $A_N^{({\rm vac})}$ becomes
\br
A_{N}^{\rho} - f_N(t)\> c \> &=&\> {\tmm}\> \varepsilon_N \> 
{{\tmm}}^{-1}\> +\> 
\partial_{N}{\tmm}\>{{\tmm}}^{-1}\> \in\> 
\bigoplus_{i\leq N_+}
\cgh_{i}(\sb)\nonu\\ 
\> &=&\> {\tpp}\> \varepsilon_N \> {{\tpp}}^{-1}\> +
\> \partial_{N} 
{\tpp}\>{{\tpp}}^{-1} \>\in\> \bigoplus_{i\geq N_-} 
\cgh_{i}(\sb) ,  
\lab{DPlus}
\er

Eqs.~\rf{tpm}, \rf{tpms2} and~\rf{DPlus} summarize the outcome of the dressing 
transformation
method, which, starting with some vacuum solution~\rf{vacpot}, associates a 
solution
of the zero-curvature equations~\rf{ZeroCurv} to each constant element $\rho$  
in the
big cell of $\GG$. The construction of this solution involves two
steps. First, the Eqs.~\rf{DPlus} can be
understood as a local change of variables between the components of the
potential $A_N$ and some components of the group elements $\tpp$  and
$\tmm$. 

The second step consists in obtaining the value of the required components
of $\tpp$  and $\tmm$ from Eqs.~\rf{tpm} and \rf{tpms2}. This is usually done 
by considering matrix elements of the form
\be
\bra{\mu}\> {{\tmm}}^{-1}\> {\tpp}  \> 
\ket{\mu^{\prime}} = 
\bra{\mu}\> e^{\sum_{N} \varepsilon_N t_N} \> \rho \> 
e^{-\sum_{N} \varepsilon_N t_N} \> \ket{\mu^{\prime}}\>,
\lab{solspec}
\ee
where $\ket{\mu}$ and $\ket{\mu^{\prime}}$ are vectors in a given 
representation 
of $\cgt$. The appropriate set of vectors is specified by the condition that   
all
the required components of ${\tmm}$  and ${\tpp}$ can  
be expressed in
terms of the resulting matrix elements. It will be shown below that the 
required matrix elements, considered as functions of the
group element $\rho$, constitute the generalization of the Hirota's            
tau-functions
for these hierarchies. Moreover, Eq.~\rf{solspec} is the analogue of the, so 
called,
solitonic specialization of the Leznov-Saveliev solution proposed
in~\ct{TUROLA,TUROLB,SOLSPEC,fms,SAVGERV} for
the  affine (abelian and non-abelian) Toda theories.

Consider now the common eigenvectors of the adjoint action of the
$\varepsilon_N$'s that specify the vacuum solution~\rf{vacpot}. Then, the
important class of multi-soliton solutions is conjectured to
correspond to group elements $\rho$ which are the product of
exponentials of eigenvectors
\be
\rho = e^{F_1} \> e^{F_2} \> \ldots e^{F_n} \>, \qquad 
[ \varepsilon_N \> , \> F_k ] = \omega_N^{(k)} \> F_k \> , \quad k=1,2, 
\ldots n\>.
\lab{eigenb}
\ee
In this case, the dependence of the solution upon the times $t_N$ can be made
quite explicit
\be
\bra{\mu}\> \tmm^{-1}\> \tpp  \> \ket{\mu^{\prime}} = 
\bra{\mu}\> \prod_{k=1}^n \exp (e^{\sum_{N} \omega_{N}^{(k)} t_N} F_k )  
 \> \ket{\mu^{\prime}}\>.
\lab{solspecb}
\ee
We emphasize that not all solutions of the type~\rf{solspecb} are soliton
solutions, but we conjecture that the soliton and multi-soliton solutions are
among them.  The conjecture that multi-soliton solutions are
associated with group elements of the form~\rf{eigenb} naturally follows from 
the
well known properties  of the multi-soliton solutions of affine Toda equations
and of hierarchies of the KdV type, and, in the sine-Gordon theory, it has been
explicitly checked in Ref.~\ct{BABT}. Actually, in all these cases, the
multi-soliton solutions are obtained in terms of representations of the ``vertex
operator'' type where the corresponding eigenvectors are nilpotent. Then, for
each eigenvector $F_k$ there exists a positive integer number $m_k$ such that
$(F_k)^{m} \not=0$ only if $m\leq m_k$. This remarkable property simplifies the
form of~\rf{solspecb} because it implies that
$e^{F_k}\> =\> 1\>+ \> F_k\> + \cdots+ (F_k)^{m_k}/m_k!$, which provides a
group-theoretical justification of Hirota's method.   
 
An interesting feature of the dressing transformations method is the
possibility of relating the solutions of different integrable equations.
Consider two different integrable hierarchies whose vacuum solutions are
compatible, in the sense that the corresponding vacuum Lax operators commute.
Then, one can consider the original integrable equations as the restriction of
a larger hierarchy of equations. Consequently, the solutions obtained through
the group of dressing transformations can also be understood in terms of the
solutions of the larger hierarchy, which implies certain relations among them.
(see section~4 of \ct{fms} for more details).

\sect{The tau-functions}

According to the discussion in the previous section, the orbits generated by
the group of dressing transformations acting on some vacuum provide
solutions of certain integrable hierarchies of equations. Making contact with
the method of Hirota, the generalized ``tau-functions'' that will be
defined in this section constitute a new set of variables to describe those
solutions. One of the characteristic properties of these variables is that they
substantially simplify the task of constructing multi-soliton
solutions~\ct{fms}. The group-theoretical interpretation of this property
has already been pointed out in the previous section. Tau-functions are given
by certain matrix elements in a appropriate representation of the Kac-Moody
Group~$\GG$. Moreover, the tau-functions corresponding to the multi-soliton
solutions are expected to involve nilpotent elements of~$\GG$, which is the
origin of their remarkable simple form. 

The tau-function formulation of the Generalized Drinfel'd-Sokolov Hierarchies
of~\ct{GEN} has already been worked out in~\ct{TAUTIM}, which, in
fact, has largely inspired our approach. However,
there are two important differences between our results and those
of~\ct{TAUTIM}. Firstly, our approach applies to the affine Toda equations
too, and, secondly, it does not rely upon the use of (level-one) vertex
operator representations. 

At this point, it is worth recalling that the
solutions constructed in sections \ref{sec:dressing} and \ref{sec:soliton} are 
completely representation-independent. 
In contrast, our definition of tau-functions makes use of a special class of
representations of the Kac-Moody algebra $\cgh$ called ``integrable
highest-weight'' representations.
The reason why these representations are called ``integrable'' is the
following. For an infinite-dimensional representation, it is generally
not possible to go from a representation of the algebra $\cgh$ to a
representation of the corresponding group $\GG$ via the exponential map
$x\mapsto {\rm e}^x$. However, the construction does work if, for
instance, the formal power series terminates at a certain power of $x$,
or if the representation space admits a basis of eigenvalues of $x$.
These conditions, applied to the Chevalley generators of $\cgh$,
single out this special type of representations. 

The generalized tau-functions will be sets of matrix elements of the form
indicated on the right-hand-side of~\rf{solspec}, considered as functions of   
the
group element $\rho$. They are characterized by the condition that they allow 
one
to parameterize all the components of $\tpp$ and $\tmm$ required to
specify the solutions~\rf{DPlus} of the zero-curvature equations~\rf{ZeroCurv}. 
As 
we
have discussed before, the tau-functions corresponding to the multi-soliton
solutions are expected to have a very simple form. However, in contrast with
the original method of Hirota, we cannot ensure in general that the
equations of the hierarchy become simpler in terms of this new set of
variables.  

First, let us discuss the generalized Hirota tau-functions associated with the
components of $B$. In equation~\rf{solspec}, these components can be isolated by
considering the vectors $\ket{\mu_0}$ of an integrable highest-weight
representation $L(\tilde\sb)$ of $\cgt$ which are annihilated by all the 
elements in $\cgh_{>0}(\sb)$, {\it i.e.\/}, $T\ket{\mu_0} = 0$ and 
$\bra{\mu_0}T'=0$ for all $T\in \cgh_{>0}(\sb)$ and $T'\in \cgh_{<0}(\sb)$,
respectively. Then, the corresponding tau-functions are defined
as~\footnote{Since the resulting  relations between tau-functions and
components of the $A_N$'s will be considered as generic changes of variables, 
we will not generally indicate the   intrinsic
dependence of the tau-functions on the group element $\rho$.}
\br
\tau_{\mu_0,\mu_{0}'}(t)\> &=&
\> \bra{\mu_{0}'}\Psi^{({\rm vac})}\> \rho\> {\Psi^{({\rm
vac})}}^{-1}\ket{\mu_{0}} \nonu\\  
&=& \> \bra{\mu_{0}'}\> e^{\sum_{N} \varepsilon_N t_N}
\>  \rho \>  e^{-\sum_{N} \varepsilon_N t_N} \> \ket{\mu_{0}}\>, 
\lab{TauB}
\er
and, in terms of them, equation~\rf{solspec} becomes just
\be
\bra{\mu_{0}'}\>  B^{-1}\> \ket{\mu_{0}}\> =\> \tau_{\mu_{0},\mu_{0}'}(t)\>.
\lab{solspecBF}
\ee
where we have denoted 
\be
B^{-1} \equiv \( \Psi^{({\rm vac})}\> \rho\> {\Psi^{({\rm vac})}}^{-1}\)_0
\ee

By construction, $\cgh_0(\sb)$ always contains the central element $C$ of the
Kac-Moody algebra, but it is always possible to split the contribution of the
corresponding field in~\rf{solspecBF}. Let $s_q\not=0$ and consider the        
subalgebra
${\cgt}^{({\rm q})}$ of
$\cgt$ generated by the $e_{i}^\pm$ with $i=0,\ldots,r$ but
$i\not=q$, which is a semisimple finite Lie algebra of rank $r$
(${\cgt}^{({\rm q})}$ is always simple if $q=0$). Then, $\cgh_0(\sb) =
\bigl(\cgh_0(\sb)\cap {\cgt}^{({\rm q})}\bigr)
\oplus {\IC}\> C$ and, correspondingly, $B$ can be split as $B= b\> \exp(\nu
\>c)$. Here, $\nu$ is the field along $C$, and $b$ is a function taking values
in the semisimple finite Lie group ${G_0}^{({\rm q})}$ whose Lie algebra is
$\cgh_0(\sb)\cap {\cgt}^{({\rm q})}$. Since
$\tilde K=\sum_{i=0}^r k_{i}^\vee \tilde s_i$ is the level of the representation
$L(\tilde\sb)$, Eq.~\rf{solspecBF} is equivalent to
\be
\bra{\mu_{0}'}\>  B^{-1}\> \ket{\mu_{0}}\> =\> {\rm
e\/}^{-\nu\> \tilde K}\>\bra{\mu_{0}'}\>  b^{-1}\> \ket{\mu_{0}}\>=\>
\tau_{\mu_{0},\mu_{0}'}(t)\>.
\lab{CentreA}
\ee
Moreover, it is always possible to introduce a tau-function for the field
$\nu$. Let us consider the highest-weight vector $\ket{v_q}$ of the
fundamental representation $L(q)$, which is obviously annihilated by all the
elements in ${\cgt}^{({\rm q})}$. Therefore,
\be
\bra{v_{q}}\>  B^{-1}\> \ket{v_q}\> =\> {\rm
e\/}^{-\nu\> k_{q}^\vee}\>= \tau_{v_q,v_q}(t)\equiv \>
\tau_{q}^{(0)}(t)\>,
\lab{CentreAlinha}
\ee
which leads to
\be
\bra{\mu_{0}'}\>  b^{-1}\> \ket{\mu_{0}}\>=\> 
{\tau_{\mu_{0},\mu_{0}'}(t) \over
\Bigl(\tau_{q}^{(0)}(t)\Bigr)^{\tilde K/k_{q}^\vee}}\>\quad  {\rm
and}\quad \nu\>= \> -\> \ln {{\tau_{q}^{(0)}(t)\over k_{q}^\vee}}\>.
\lab{CentreB}
\ee

Finally, recall that the vectors $\ket{\mu_0}$ form a representation of
the semisimple Lie group ${G_0}^{({\rm q})}$. Therefore, if $L(\tilde\sb)$ is
chosen such that this representation is faithful, Eq.~\rf{CentreB} allows one  
to
obtain all the components of $b$ in terms of the generalized tau-functions
$\tau_{\mu_{0},\mu_{0}'}$ and $\tau_{q}^{(0)}$. Notice 
that, in this case, the definition of generalized tau-functions coincide 
exactly with the quantities involved in the solitonic specialization of the
Leznov-Saveliev solution proposed in~\ct{SOLSPEC}. 

Let us now discuss the generalized tau-functions associated with the components
of $\tms$. Consider the gradation $\sb$ of $\cgt$ involved in the
definition of the integrable hierarchy. For each $s_i\not=0$, let us consider
the highest-weight vector of the fundamental representation $L(i)$ and define
the (right) tau-function vector
\br
\ket{\tau_{i}^R(t)}\> & =\> \Psi^{({\rm vac})}\> h\> {\Psi^{({\rm
vac})}}^{-1}\> \ket{v_i} \nonu\\  
& = \>  e^{\sum_{N} \varepsilon_N t_N} \> h \> 
e^{-\sum_{N} \varepsilon_N t_N} \> \ket{v_i}
\lab{TauFunc}
\er
Notice that $\ket{\tau_{i}^R(t)}$ is a vector in the representation $L(i)$.
Therefore, it has infinite components, and it will be shown soon that the role
of the Hirota tau-functions will be played by a finite subset of them.
Taking into account that $\ket{v_i}$ is annihilated by all the elements in
$g_{>0}(\sb)$, equation~\rf{solspec} implies
\be
{\tms}^{-1}\> B^{-1}\> \ket{v_i}\> =\> \ket{\tau_{i}^R(t)} \>,\quad
i=0,\ldots,r \quad{\rm and}\quad s_i\not=0\>.
\lab{SolspecT}
\ee

The definition~\rf{TauFunc} is inspired by the tau-function approach
of~\ct{KW,TAUTIM,fms}. However,
in~\ct{TAUTIM}, and~\ct{fms}, the authors consider a unique tau-function
$\ket{\tau_{\sb}(t)} \in L(\sb)$. In fact, one could equally consider different
tau-functions $\ket{\tau_{\sb'}(t)}$ associated with any integrable
representation $L(\sb')$ such that $s_{i}'\not=0$ if, and only if,
$s_i\not=0$. Since the highest weight state of $L(\sb)$ is obtained by the
tensor product of the $\ket{v_i}$'s, as 
\be
\ket{v_\sb} \>=\> \bigotimes_{i=0}^{r}\> \bigl\{ \ket{v_i}^{\otimes s_i}
\bigr\}\>.
\lab{Tensor}
\ee
one sees that all these choices lead to the same results, but the one presented
here is the most economical.  

Since, for any integrable representation, the grading operator  \rf{gradop}    
can be
diagonalized acting on $L(\sb)$, these tau-functions vectors can be decomposed 
as
\be
\ket{\tau_{i}^R (t)}\> =\> \sum_{-j\in {\IZ}\leq0} \ket{\tau_{i}^{R(-j)}
(t)}\>,
\qquad Q_i \> \ket{\tau_{i}^{R(-j)} (t)}\> =\> -j \>\ket{\tau_{i}^{R(-j)}
(t)}\>, 
\lab{CTau}
\ee
where we have used that $\tms\in \GG_{<0}(\sb)$ and $B\in \GG_0(\sb)$, and
$Q_i$ indicates the derivation corresponding to the grading operator \rf{gradop}
 with $s_j=\delta_{j,i}$. Moreover, the highest-weight vector is  
an eigenvector of the subalgebra $\cgh_0(\sb)$ and, consequently, of $B$. 
Therefore,
\be
\ket{\tau_{i}^{R(0)} (t)}\> =\> B^{-1}\> \ket{v_i}\> =\>
\tau_{i}^{(0)} (t) \> \ket{v_i}\>,
\lab{EienZero}
\ee
where, $\tau_{i}^{(0)} (t)$  is a ${\IC}$-function, not
a vector of $L(i)$, whose definition is~\footnote{To compare
with~\rf{solspecBF}, notice that $\ket{\mu_0} = \ket{v_i}$ forms a 
one-dimensional
representation of $\cgh_0({\sb})$ and, consequently, $\tau_{v_i,\mu_{0}'}(t)$
vanishes unless
$\ket{\mu_{0}'}=\ket{v_i}$. Therefore, for non-abelian $G_0$, the
required tau-functions $\tau_{\mu_0,\mu_{0}'}(t)$ have to involve the
fundamental integrable representations $L(j)$ corresponding to $s_j=0$, in
contrast with $\ket{\tau_{i}^R(t)}$ (see Eq.~\rf{SolspecT}).}   
\be
\tau_{i}^{(0)} (t) \>=\> 
\bra{v_i}\> e^{\sum_{N} \varepsilon_N t_N}
\> \rho\>\>
e^{-\sum_{N} \varepsilon_N t_N}\> \ket{v_i}\>\equiv \> \tau_{v_i,v_i}(t)
\lab{TauCero}
\ee
(compare with Eq.~\rf{CentreA}). Therefore, Eq.~\rf{SolspecT} becomes
\be
{\tms}^{-1}\> \ket{v_i} \> =\> {1\over \tau_{i}^{(0)} (t)}\> \ket{\tau_{i}^R
(t)}\>,
\lab{TetaTau}
\ee
which is the generalization of the Eq.~(5.1) of~\ct{TAUTIM} for general
integrable highest-weight representations of $\cgt$. Eq.~\rf{TetaTau}
allows one to express all the components of $\tms$ in terms of
the components of $\ket{\tau_{i}^R (t)}$ for all $i=0,\ldots,r$ with
$s_i\not=0$ (for instance, by using the positive definite Hermitian form of
$L(i)$). However, it is obvious that only a finite subset of them enter in the
definition of the potentials $A_N$ through Eq.~\rf{DPlus}.

In exactly the same way, one can introduce another set of ``left''
tau-function vectors through
\be
\bra{\tau_{i}^L(t)}\>  =\> \bra{v_i}\> \Psi^{({\rm vac})}\> h\> {\Psi^{({\rm
vac})}}^{-1}\>,
\ee
which leads to
\be
\bra{v_i}\> \tpg \> =\> \bra{\tau_{i}^L (t)}\> {1\over
\tau_{i}^{(0)} (t)}\>,
\ee
and allows one to express all the components of $\tpg$ in terms of the
components of $\bra{\tau_{i}^L (t)}$ for all $i=0,\ldots,r$ with
$s_i\not=0$.

Summarising, the generalized Hirota tau-functions of these hierarchies consist
of the subset of functions $\tau_{\mu_0,\mu_{0}'}$ and of components of
$\ket{\tau_{i}^{R}}$ and $\bra{\tau_{i}^{L}}$ required to parameterize all the
components of the potentials $A_N$ in Eq.~\rf{DPlus}. Then, for the            
multi-soliton
solutions corresponding to the group element $\rho$ specified in~\rf{eigenb},  
their
truncated power series expansion follows from the
possible  nilpotency of the eigenvectors $F_k$ in these representations. For
instance, if
$n=1$ in~\rf{eigenb} and 
$F_{1}^m\ket{\mu_0} =F_{1}^m\ket{v_i} =0$ unless $m\leq m_1$, then 
\br
\tau_{\mu_0,\mu_{0}'}(t)\>& = &\> \tilde\tau_{\mu_0,\mu_{0}'}^{0}\>
+\> 
\tilde\tau_{\mu_0,\mu_{0}'}^{1} \>+\> 
\ldots + \tilde\tau_{\mu_0,\mu_{0}'}^{m_1} \nonu\\ 
& = &\sum_{k=0}^{m_1}\> {1\over k!\>}\> {\rm e\>}^{k\> \sum_{N} \> \omega_N
t_N}\>\bra{\mu_{0}'} \>
F_{1}^{k}\> \ket{\mu_0}\>,\quad {\rm and}\nonu\\   
\ket{\tau_{i}^R(t)} \>& = &\> \sum_{k=0}^{m_1}\> {1\over k!\>}\> {\rm e\>}^{k\> 
\sum_{N} \> \omega_N t_N}\>F_{1}^{k}\> \ket{v_i}\>. 
\lab{trunca2}
\er

\sect{The example of the Non Abelian To\-da \\
Mo\-dels}
\label{sec:formulation}

Consider an untwisted affine Kac-Moody algebra $\cgh$ endowed with an integral 
gradation $\cgh = \bigoplus_{n\in \IZ} \cgh_n$ (see
\rf{Gradation},\rf{gradop}).       
By an affine Kac-Moody algebra we mean a loop algebra corresponding to a
finite dimensional simple Lie algebra $\cg$ of rank $r$, extended by the center
$C$ and the derivation $D$.

Let ${\cal M}$ be a two dimensional manifold with local coordinates 
$x_+$ and $x_-$; $\hat{\cal G}$ be an affine Kac-Moody algebra corresponding
to a finite dimensional complex simple Lie algebra ${\cal G}$ with 
the Lie group $G$; ${\cal A}$ be a flat connection in the trivial holomorphic 
principal fibre bundle ${\cal M}\times \hat{G}\longmapsto {\cal M}$.
Specify the connection in such a way that its $(1,0)$-component takes values in
the subspaces $\bigoplus_{n=0}^{l} \cgh_{+n}$, and $(0,1)$-component takes
values in  $\bigoplus_{n=0}^{l} \cgh_{-n}$, with $l$ being a fixed positive
integer. In other words, up to a relevant gauge tranformation, these
components, satisfying the zero curvature condition
\be
\pa_{+} A_{-} - \pa_{-} A_{+} + \lb A_{+}\, , \, A_{-} \rb = 0,
\lab{zc}
\ee
are of the form
\br
A_{+} = - B\, F^{+} \, B^{-1} \, , \qquad
A_{-} = - \pa_{-} B \,  B^{-1} + F^{-}.
\lab{gp}
\er
Here $B$ is a mapping from ${\cal M}$ to the Lie group $\hat{G}_0$ with
the Lie algebra  $\cgh_0$; $F^{\pm}$ are mappings to $\bigoplus_{n=1}^{l}
\cgh_{\pm n}$ of the form
\be
 F^{+} = E_{l} + \sum_{m=1}^{l-1} F^{+}_m \, , 
\lab{fp}\,\,  \qquad 
 F^{-} = E_{-l} + \sum_{m=1}^{l-1} F^{-}_m\, ,
\ee
with $E_{\pm l}$ being some fixed elements of $\cgh_{\pm l}$; and $F^{\pm}_m$,
$1\leq m\leq l-1$, take values in $\cgh_{\pm m}$.  

Substituting the gauge potentials \rf{gp} into \rf{zc}, one gets the equations 
of motion
\br
\pa_{+}\( \pa_{-} B\, B^{-1}\) &=& \lb E_{-l} \, , \, B \, E_{l}\, B^{-1}\rb 
+ \sum_{n=1}^{l-1} \lb F^{-}_{n} \, , \, B \, F^{+}_{n}\, B^{-1}\rb \, ,
\lab{em1}\\
\pa_{-} F^{+}_{m} &=& \lb E_{l} \, , \, B^{-1} \, F^{-}_{l-m}\,B \rb 
+ \sum_{n=1}^{l-m-1} \lb F^{+}_{n+m} \, , \, B^{-1} \, F^{-}_{n}\, B\rb \, ,
\lab{em2}\\
\pa_{+} F^{-}_{m} &=& -\lb E_{-l} \, , \, B \,  F^{+}_{l-m}\,B^{-1} \rb  
- \sum_{n=1}^{l-m-1} \lb F^{-}_{n+m} \, , \, B \, F^{+}_{n}\, B^{-1}\rb \, .
\lab{em3}
\er

Since $Q_{{\bf s}}$, defined in \rf{gradop}, and $C$ are in $\cgh_0$, we       
parametrise $B$ as
\be
B = b\, e^{\eta \, Q_{{\bf s}}} \, e^{\nu \, C}\, ,
\lab{bdef}
\ee
where $b$ is a mapping to $G_0$, the subgroup of $\hat{G}_0$ generated by all
elements of $\cgh_0$ other than $Q_{{\bf s}}$ and $C$. The fields $\eta$ and
$\nu$ correspond to the extension of the loop algebra, and, as we will show
below, are responsible for making the system conformally invariant
\ct{AFGZ,bb}. Clearly, the order of the three factors in \rf{bdef} is
irrelevant, since they commute. In addition, we will use a special basis for
the generators of $\cgh_0$ such that they are all orthogonal to 
$Q_{{\bf s}}$ and 
$C$. From \rf{gradop} one observes that the generators of $\cgh_0$ are, besides
$C$ and $Q_{{\bf s}}$, the elements $H^0_a$, $a=1,2,\ldots r$, of the
 Cartan subalgebra, 
and step operators $E_{\pm\a}^0$ and $E_{\pm\b}^{\mp 1}$, such that
$\sum_{a=1}^{r} s_a \l^v_a \cdot \a =0$, and 
$\sum_{a=1}^{r} s_a \l^v_a \cdot \b = N_{{\bf s}}$. 
There can be no step operators $E_{\gamma}^{n}$, with $\mid n
\mid >1$, as explained in appendix C of ref. \ct{fms}. Therefore, 
shifting the Cartan elements as
\br
\widetilde{H}_{a}^0\, =\, H_{a}^0\, -\, {1\over N_{\bf s}}\, \Tr\left(H_{\bf
s}\, H_{a}^0\right)\, C =\, H_{a}^0\, -\, {2\over
\alpha_{a}^2}{s_a\over N_{\bf s}}\, C,
\lab{cartanplus}
\er 
one gets
\br &&\Tr\left(C^2\right)= \Tr\left(C\,\widetilde{H}_{a}^0\right) = \Tr\left(
Q_{\bf s}^2 \right)=
\Tr\left( Q_{\bf s}\, \widetilde{H}_{a}^0\right)=0,
 \quad 
\Tr\left( Q_{\bf s}\, C\right) = N_{\bf s},\nonu \\
&&\Tr\left(
\widetilde{H}_{a}^0\, \widetilde{H}_{b}^0\right)= \Tr\left(H_{a}^0\,
H_{b}^0\right) = 4\a_a \cdot \a_b/\a_a^2 \a_b^2\equiv \eta_{ab},
\lab{bilbasis1}
\er 
for all $a,b=1\ldots,r$. \\
Here we have used $H_a^0= 2 \a_a\cdot H^0/\a_a^2$, 
$\Tr\( x\cdot H^0\, y\cdot H^0\) = x \cdot y$, and $\Tr \( C\,D\) =1$. For more
detail of such a special basis, see appendix C of ref. \ct{fms}.

Substituting \rf{bdef} into the equations of motion \rf{em1}--\rf{em3}, 
one has
\br
\pa_{+} \( \pa_{-} b b^{-1}\) &+& 
\pa_{+}\pa_{-} \, \nu \, C = e^{l\eta} \lb E_{-l} \, , \, b\, E_{l}\,
b^{-1}\rb + \sum_{n=1}^{l-1} e^{n\eta}\, \lb F^{-}_n\, ,\, b\,
F^{+}_{n}b^{-1}\rb 
\lab{eqm1}\\
\pa_{-} F^{+}_m &=& e^{(l-m)\eta}\, \lb E_{l} \, , \, b^{-1}\, F^{-}_{l-m}\, b
\rb + \sum_{n=1}^{l-m-1} e^{n\eta} \lb F^{+}_{m+n} \, , \, b^{-1}\, F^{-}_n\,
b\rb \, ,
\lab{eqm2}\\
\pa_{+} F^{-}_m &=& -e^{(l-m)\eta}\, \lb E_{-l} \, , \, b\, F^{+}_{l-m}\,
b^{-1} \rb 
- \sum_{n=1}^{l-m-1} e^{n\eta} \lb F^{-}_{m+n} \, , \, b\, F^{+}_n\, b^{-1}\rb
\, ,
\lab{eqm3}\\
\pa_{+}\pa_{-}\, \eta \,  Q_{{\bf s}}  &=& 0 \, ,
\lab{eqm4} 
\er
where the last equation is a consenquence of the fact that $D$, and hence
 $Q_{{\bf
s}}$, can not be obtained as the Lie bracket of any two elements of $\cgh$.

The structure of the vacuum of the system \rf{eqm1}--\rf{eqm4} is rather
complicated. We will discuss some aspects of it below. However, there is a
simple condition that guarantees the existence of static (vacuum) solutions. 
If the elements $E_{\pm l}$ satisfy the relation
\be
\lb E_{l}\, , \, E_{-l}\rb = \b \, C \, , \qquad 
\mbox{\rm where \quad  $\b = {l\o  N_{\bf s}}\, \Tr \( E_{l}\, E_{-l}\)$},
\lab{vaccond}
\ee
then 
\be 
b=1 \, , \qquad F^{\pm}_{m} = 0 \, , \qquad \eta = 0 \, , \qquad 
\nu = -\b x_{+} x_{-},
\lab{vacuum1}
\ee
is a (vacuum) solution of \rf{eqm1}--\rf{eqm4}. 
  
Another possibility for vacuum solutions arises when $E_{\pm l}$, $l>1$, belong
to a Heisenberg subalgebra of $\cgh$, see \ct{kac1,kacpet},
\be
\lb E_{M}\, , \, E_{N}\rb = \Tr \( E_{M}E_{-M}\) \, M \, \d_{M+N,0}\, C,
\lab{heis}
\ee  
where $M, N$ belong to some (infinite) subset $\IZ_E$ of the integer numbers
$\IZ$. In such cases one has that  
\br  
b=1 \, , \quad 
\eta = 0 \, , \;  F^{\pm}_M =c^{\pm}_M\, E_{\pm M} \, , \; 
F^{\pm}_m = 0 \, , \; \mbox {\rm  $m \notin \IZ_E$}\, , \; 
\nu = - \Omega \, x_{+}\, x_{-}, 
\lab{vacuum2}
\er 
is a solution of \rf{eqm1}--\rf{eqm4} with $c^{\pm}_M$ being constants, and
\be
\Omega \equiv  \b + \sum_{M=1}^{l-1} \Tr\( E_{M}E_{-M}\)\,  M\, c^{+}_M\,
c^{-}_M.
\lab{omega} 
\ee

Obviously, the system \rf{eqm1}--\rf{eqm4} may have many more vacuum solutions
besides \rf{vacuum1} and \rf{vacuum2}. However,  the condition \rf{vaccond}
guarantees the existence of at least one vacuum solution. Such a fact, as we 
will see below, favors the existence of soliton solutions.

The models introduced above are completely characterised by the data   
$\{\cgh , Q_{{\bf s}}, l, E_{\pm l}\}$; and we have a quite large class of 
systems with physical properties crucially depending on a choice of 
those data. 

Equations \rf{eqm1} -- \rf{eqm4} are invariant under the conformal
transformation
\be
x_{+} \ra f(x_{+}) \, , \qquad x_{-} \ra g(x_{-}),
\lab{ct}
\ee
with $f$ and $g$ being analytic functions; and with the fields transforming as
\br
b(x_{+}\, , \, x_{-}) &\ra& 
{\tilde b}({\tilde x}_{+}\, , \,  {\tilde x}_{-}) = b(x_{+}\, , \, x_{-}) \, ,
\lab{ctf1}\\
e^{-\nu (x_{+}\, , \, x_{-})} &\ra& e^{-{\tilde \nu}({\tilde x}_{+}\, , \, 
{\tilde x}_{-})} = \( f^{\pr}\)^{\d} \, \( g^{\pr}\)^{\d}
e^{-\nu (x_{+}\, , \, x_{-})} \, ,
\lab{ctf2}\\
e^{-\eta (x_{+}\, , \, x_{-})} &\ra& e^{-{\tilde \eta}({\tilde x}_{+}\, , \, 
{\tilde x}_{-})} = \( f^{\pr}\)^{1/l} \, \( g^{\pr}\)^{1/l}  e^{-\eta (x_{+}\, 
, \, x_{-})} \, ,
\lab{ctf3}\\ 
F^{+}_m (x_{+}\, , \, x_{-}) &\ra & {\tilde F}^{+}_m ({\tilde x}_{+}\, , \, 
{\tilde x}_{-}) =   \( f^{\pr}\)^{-1+m/l}\, F^{+}_m (x_{+}\, , \, x_{-}) \, ,
\lab{ctf4}\\
F^{-}_m (x_{+}\, , \, x_{-}) &\ra & {\tilde F}^{-}_m ({\tilde x}_{+}\, , \, 
{\tilde x}_{-}) =   \( g^{\pr}\)^{-1+m/l}\, F^{-}_m (x_{+}\, , \, x_{-}) \, ,
\lab{ctf5}
\er
where the conformal weight $\d$, associated to $e^{-\nu}$, is arbitrary.

Notice that the Lorentz transformation $ x_{\pm} \ra \l^{\mp 1} x_{\pm}$ is 
obtained from \rf{ct} by taking $f(x_{+})=x_{+}/\l$ and $g(x_{-}) = \l x_{-}$.

Equations \rf{eqm1}--\rf{eqm4} are also invariant under the 
transformations 
\br
b\( x_{+}\, ,\, x_{-}\) &\ra& h_L\(x_{-}\)\, b\( x_{+}\, ,\, x_{-}\)\, 
h_R\(x_{+}\),
\lab{gs1}\\
F^{+}_m\( x_{+}\, ,\, x_{-}\) &\ra& h_R^{-1}\(x_{+}\)\, F^{+}_m\( x_{+}\, ,\,
x_{-}\)\,  h_R\(x_{+}\),
\lab{gs2}\\
F^{-}_m\( x_{+}\, ,\, x_{-}\) &\ra& h_L\(x_{-}\)\, F^{-}_m\( x_{+}\, ,\,
x_{-}\)\,  h_L^{-1}\(x_{-}\),
\lab{gs3}
\er
where $h_L\(x_{-}\)$ and $h_R\(x_{+}\)$ are elements of subgroups
$\ch_0^L$ and $\ch_0^R$ of $G_0$, respectively, satisfying the conditions 
\br
h_R\(x_{+}\)\, E_{l} \, h_R^{-1}\(x_{+}\) = E_{l}\, , \qquad 
h_L^{-1}\(x_{-}\)\, E_{-l} \, h_L\(x_{-}\) = E_{-l}.
\lab{hlr}
\er
The left and right gauge transformations commute, and so the gauge group is 
$\ch_0^L \otimes \ch_0^R$. Whenever $\ch_0^L$ and $\ch_0^R$ have a set of
common generators, we get an important subgroup of the gauge group, namely
$\ch_D \equiv \ch_0^L \cap \ch_0^R$. These are global gauge transformations,
where the fields are transformed under conjugation ($h_L=h_R^{-1}\equiv h_D 
= {\rm const.}$), 
\be
b \ra h_D\, b h_D^{-1} \, , \qquad F^{\pm}_m \ra h_D\, F^{\pm}_m h_D^{-1}, 
\lab{diagonalgauge}
\ee
and $E_{\pm l} = h_D\, E_{\pm l} h_D^{-1}$. We discuss the relevance
of these transformations below.
 
\sect{Soliton Solutions}
\label{sec:solitontoda}

We now perform the dressing transformation, dicussed in sections
\ref{sec:dressing} and \ref{sec:soliton}, by taking as an initial
configuration a vacuum solution of \rf{eqm1}--\rf{eqm4}. As we have said, the
model under consideration may have several type of vacuum solutions. However,
here we will deal with the solutions of type \rf{vacuum1} or \rf{vacuum2}. 

For the vacuum solutions \rf{vacuum2}, the gauge potentials \rf{gp} become
\br  
A_{+}^{(0)} = - \ce_{+}\, , \qquad A_{-}^{(0)} =  \ce_{-} + \Omega x_{+} C,
\lab{vacgp}
\er  
with ${\cal E}_{\pm}$ given by 
\be
\ce_{\pm} \equiv E_{\pm l} + \sum_{N=1}^{l-1} c^{\pm}_N E_{\pm N}\, , \quad
\mbox{\rm and so} \quad 
\lb \ce_{+} \, , \, \ce_{-} \rb = \Omega \, C;
\lab{cepm}
\ee 
where $c^{\pm}_N$ and $\Omega$ were introduced in \rf{vacuum2} and \rf{omega},
respectively. 

They can be written as
\be A_{\pm}^{(0)} = - \pa_{\pm} \Psi^{({\rm vac})} \, 
{\Psi^{({\rm vac})}}^{-1}, \quad \mbox{\rm with} \quad 
 \Psi^{({\rm vac})} = e^{x_{+}\, \ce_{+}}\, e^{-x_{-}\, \ce_{-}}.
\lab{t0}
\ee

The gauge potentials for the vacuum solution \rf{vacuum1} are
obtained from \rf{vacgp} by taking $c^{\pm}_n =0$. In fact, they are connected
by the gauge transformation 
\be
A_{\pm}^{(0)} = {\tilde {\Psi}}^{({\rm vac})}
A_{\pm}^{(0)}\mid_{c^{\pm}_n=0}\, \({\tilde \Psi}^{({\rm vac})}\)^{-1} - 
\pa_{\pm} {\tilde {\Psi}}^{({\rm vac})}
\({{\tilde \Psi}^{({\rm vac})}}\)^{-1}, 
\ee
with 
\be
 {\tilde {\Psi}}^{({\rm vac})} 
= \exp [ x_{+} \(\ce_{+} - 
E_{l}\) ] \exp [ -x_{-} \(\ce_{-} - E_{-l}\) ].
\ee  
However, in general, the vacuum solutions \rf{vacuum1}
and \rf{vacuum2} may not be connected by any dressing transformation,
and, in such a case, the
existence of two elements of  form \rf{tpm}, is not always possible. 
Consequently, one can have soliton solutions lying on different
orbits under the dressing transformations. 

In order to perform the dressing procedure, we  take \rf{vacgp} 
as  initial gauge potentials. 
Then, we obtain, under the dressing procedure, the solutions on 
the orbit of vacuum \rf{vacuum2}, and for $c^{\pm}_n=0$ those on the orbit
of the vacuum  \rf{vacuum1}.  From the structure of the dressing
transformations and from the fact that the grading operator
\rf{gradop} is never the result of any commutation, since it contains $D$, it
follows that the dressing transformations do not excite the field $\eta$. 
Therefore, from  \rf{gp}, \rf{bdef}, \rf{vacgp} and  \rf{DPlus} we get
\be 
b\, E_{l} \, b^{-1} + \sum_{m=1}^{l-1} b\, F^{+}_{m} \, b^{-1} = 
\Theta_{\pm} \( E_{l}  + \sum_{n=1}^{l-1} c^{+}_{n}
\, E_{n}   + \, \Theta_{\pm}^{-1}\, \pa_{+} \Theta_{\pm}\) 
\Theta_{\pm}^{-1},
\lab{work1}
\ee
\br
 -\pa_{-}\, b b^{-1} - \( \pa_{-} \, \nu +  \Omega x_{+} \)
\, C &+& E_{-l} +
\sum_{m=1}^{l-1}  F^{-}_{m} = \\  
&=&\Theta_{\pm}\( E_{-l}  + \sum_{m=1}^{l-1} c^{-}_{m}
\, E_{-m}  -\Theta_{\pm}^{-1}\, \pa_{-} \Theta_{\pm}\) 
\Theta_{\pm}^{-1}. \nonu 
\lab{work2}
\er
Note that in the above relations, the fields $b$, $\nu$ and
$F^{\pm}_{m}$ stand  for the solutions on the orbit of the vacuum solution
\rf{vacuum2}. The procedure to construct the solution requires  to split the
above equations into the eigensubspaces of the grading operator \rf{gradop}.
It is convenient to write
\br
\tpg = \exp \( \sum_{s>0} t^{(s)}\) \, , \qquad
\tms = \exp \( \sum_{s>0} t^{(-s)}\) \, , \mbox{ where }
\; t^{(\pm s)} \in \cgh_{\pm s}.
\lab{tpms}
\er
The mappings $t^{(\pm s)}$, for each choice of $\rho$, are determined from
\rf{tpms2} with $\Psi$ being $\Psi^{({\rm vac})}$ given in \rf{t0}. Then, the   
components of
\rf{work1} and \rf{work2} in each eigensubspace, give an equation connecting
the fields with $t^{(\pm s)}$. Thus the solutions for the  fields $b$, $\nu$ 
and $F^{\pm}_{m}$ are determined from $t^{(\pm s)}$. Such a procedure is rather
cumbersome, but fortunately, one needs to know very few $t^{(\pm s)}$'s to get
the solution. For instance, taking relations \rf{work1} and \rf{work2} for
$\Theta_{+}$ ($\Theta_{-}$) with grade components $0$ and $-l$ ($l$ and $0$),
one gets 
\be 
\tp0 = h_L^{-1} (x_{-})
\, , \qquad 
\tm0 = b\, e^{\( \nu + \Omega x_{+} x_{-}\)\, C}\, h_R (x_{+}),
\lab{tpm0}
\ee 
with $h_L (x_{-})$ and $h_R (x_{+})$ defined in \rf{hlr}.

{}From \rf{Factor}, \rf{tpm}, \rf{t0} and \rf{tpm0}  it follows that
\be 
{\tms}^{-1}\, \( h_L(x_{-})\, b\, e^{\( \nu + \Omega x_{+} x_{-}\)\, 
C}\, h_R(x_{+})\)^{-1}\, \tpg =  e^{x_{+}\, \ce_{+}} \, e^{-x_{-}\, \ce_{-}} \,
\rho \, e^{x_{-}\, \ce_{-}}\, e^{-x_{+}\, \ce_{+}}.
\lab{nice}
\ee

The space--time dependence of the r.h.s. of the above relation is given
explicitly. One can extract the solutions out of \rf{nice} by taking the
expectation  value of its both sides between suitable states of a given
representation of $\cgh$, in a  similar way to that one explained in section
\ref{sec:soliton}. 

The solitons solutions are obtained from \rf{nice} by choosing the 
fixed group element $\rho$, characterising the dressing
transformation, as the exponential of an eigenvector of $\ce_{\pm}$, i.e.
\be
\rho = e^V.
\ee 
That is the {\em solitonic specialization} discussed in section
\ref{sec:soliton}. Indeed, if $V$ satisfies the relations
\be
\lb \ce_{\pm} \, , \, V \rb = \omega_{\pm} \, V,
\ee
then \rf{nice} reads as
\be
\exp \( e^{x_{+}\, \omega_{+} - x_{-}\,
\omega_{-}} \, V\) \equiv
\exp \( e^{\gamma \( x - vt\)}\, V\) ,
\lab{nicesoliton}
\ee
with $\gamma = \omega_{+} + \omega_{-}$, and $v = \( \omega_{-} - \omega_{+}\)/
\( \omega_{+} + \omega_{-}\)$, since $x_{\pm} = t \pm x$. 

Therefore, for each eigenvector $V$, expression \rf{nicesoliton} corresponds 
to a solution 
that travels with a constant velocity $v$ without dispersion.  Depending upon
the properties of $V$, as we will see below in the examples, such solutions
correspond to one--soliton solutions.

The multi--soliton solutions are obtained by taking $\rho$ to be the product of
several one--soliton $\rho$'s, i.e.,
\be
\rho = e^{V_1}\, e^{V_2}\, e^{V_3}\, \ldots e^{V_N},
\ee
with each $V_i$ satisfying $\lb \ce_{\pm} \, , \, V_i \rb = 
\omega_{\pm}^i \, V_i$.

Notice that, under the global gauge transformations \rf{diagonalgauge}, the
gauge potentials \rf{gp} are transformed as $A_{\pm} \ra h_D \, A_{\pm} \,
h_D^{-1}$. Therefore, since the potencials are pure gauge, 
$A_{\mu} = - \pa_{\mu} T T^{-1}$, one has $T \ra h_D T$, and
consequently \rf{tpms2} implies $\Theta_{+}^{>} \ra h_D\, \Theta_{+}^{>} \,
h_D^{-1}$ and $\Theta_{-}^{<} \ra h_D\, \Theta_{-}^{<} \, h_D^{-1}$.
Hence, with solution \rf{nice} corresponding to a fixed element
$\rho$, a solution, obtained from that by a global gauge
transformation \rf{diagonalgauge}, is given by \rf{nice} with the replacement
\be
\rho \ra  h_D\, \rho \, h_D^{-1},
\lab{symrho}
\ee
if the condition $h_D\, \ce_{\pm} \, h_D^{-1} = \ce_{\pm}$ is satisfied.
For the solutions on the orbit of the vacuum \rf{vacuum1}, that is indeed 
true, since $\ce_{\pm} = E_{\pm l}$; see \rf{cepm}. For the solitonic case, 
one then obtains for each eigenvector $V$ of $\ce_{\pm}$, an orbit
of equivalent one--soliton (or multi--soliton) solutions generated by $h_D\, V
\, h_D^{-1}$.

\sect{Masses of fundamental particles and\\
 solitons}
\label{sec:masses}

As we have seen above, the system under consideration is conformally invariant.
Therefore, since we do not have a continuum mass spectrum, its fundamental
particles have to be massless.  However, such a symmetry can be spontaneoulsy
broken by choosing a particular constant solution for the field
$\eta$, say $\eta=\eta_0$. The resulting theory is then massive.
Representing the mapping $B$ as $B \equiv \exp T$, and  
considering only the linear  field approximation, i.e., the free part
of the equations of motion  \rf{em1}--\rf{em3}, one gets 
\br
\pa_{+}\pa_{-} \, T &=& - v_{\eta}\lb E_{-l} \, , \, \lb E_{l}\, , \, T \rb \rb
\, ,
\lab{mass1}\\
\pa_{+}\pa_{-} F^{+}_{m} &=& - v_{\eta}\lb E_{-l} \, , \, \lb E_{l}\, , \,
F^{+}_{m}\rb \rb \, ,
\lab{mass2}\\
\pa_{+}\pa_{-} F^{-}_{m} &=& - v_{\eta} \lb E_{-l} \, , \, \lb E_{l}\, , \,
F^{-}_{m}\rb \rb \, ,
\lab{mass3}
\er
where $v_{\eta} = e^{l\, \eta_0}$. 

Therefore, the masses of  fundamental particles in such a theory are given by
the eigenvalues of the operator $\lb E_{-l}\, , \, \lb E_{l}\, , \, * \rb
\rb$  in the subspaces $\cgh_n$, $n=0,\pm 1, \pm 2,\ldots \pm (l-1)$, i.e.,
\be
\lb E_{-l} \, , \, \lb E_{l}\, , \, X \rb \rb = \l \, X.
\lab{fundeigen}
\ee
Since $\pa_{+}\pa_{-} = {1\o 4}(\pa_t^2 - \pa_x^2)$, we obtain 
the masses from the Klein--Gordon type equations \rf{mass1} -- \rf{mass3} as  
\be
m_{\l}^2 = 4\,\l \, v_{\eta}.
\lab{massf}
\ee
That result constitute a generalization of the arguments used in the abelian 
and non abelian affine Toda models \ct{fring,fms}. 
Of course, we are interested in those cases where the eigenvalues of the
operator $\lb E_{-l}\, , \, \lb E_{l}\, , \, * \rb \rb$ are real and
positive on the subspaces under consideration.  That will be, in fact, one of 
the conditions we use to select the data $\{ \cgh , Q_{{\bf s}}, l, E_{\pm
l}\}$ for defining  physical models through \rf{gp}. 

Notice that the field $e^{l\,\eta}$ plays the role of a Higgs field, since it
not only spontaneously breaks  the conformal symmetry, but also because its
vacuum expectation value sets the mass scale of the theory. We have here the
same mechanism as in non abelian affine Toda theories \ct{acfgz,fms}.

Let us explain now, following the reasonings of \ct{acfgz}
and \ct{fms}, that
the masses of solitons are also generated by the spontaneous breakdown of
the conformal symmetry.  

The energy momentum tensor of such theories is of the form (see \ct{SAVGERV}
for more details)
\be
L_{\mu\nu}=  \Theta_{\mu\nu}  + S_{\mu\nu}\, .
\lab{teta}
\ee
where $S_{\mu\nu}$ is the improvement term
\br
S_{\mu\nu} &\equiv& -{k\o l} \Tr \( Q_{\bf s} \( \pa_{\mu}\, \( B^{-1}
\pa_{\nu} \, B \) - g_{\mu\nu} \pa_{\rho} \( B^{-1} \pa^{\rho} \, B \) \)\)
\nonu\\
&=& - {k\, N_{\bf s}\o l} \( \pa_{\mu} \pa_{\nu} -  g_{\mu\nu} \pa^2\) \nu ,
\lab{smunu}
\er
Due to the fact we are dealing with a conformally invariant theory,
$L_{\mu\nu}$ satisfies
\be
\pa_{-}\, L_{++}=0 \, , \qquad  \pa_{+}\, L_{--}=0 \, ,
\qquad L_{+-} = L_{-+} =0 \, .
\lab{conserve}
\ee
Even though it is not traceless, $\Theta_{\mu\nu}$ is 
symmetric and conserved,
\be
\pa^{\mu}\,\Theta_{\mu\nu}=0 \, ,
\lab{tetacons} 
\ee

The energy of  classical solutions are given by the
space integral of the $(0,0)$ component of energy--momentum tensor
$L_{\mu\nu}$. In the Lorentz frame 
where the classical soliton  solution is static, the energy should be 
interpreted as the mass of the soliton. However, since the  theory is 
conformally 
invariant, it has no mass scale, and the soliton mass  should vanish. When the
conformal symmetry is spontaneously broken by choosing  a particular constant
solution for the field $\eta$,  we obtain a massive theory. Construct the
energy--momentum tensor of such a theory as follows. Clearly, the tensor
$\Theta_{\mu\nu}$, introduced in \rf{teta} and evaluated at any classical
solution, satisfies \rf{tetacons}. Therefore, the tensor defined by 
\be
\Theta^{\rm broken}_{\mu\nu} \equiv \Theta_{\mu\nu}\mid_{\eta ={\rm constant}}
\, ,
\ee
is symmetric and conserved,
\be
\pa^{\mu}  \Theta^{\rm broken}_{\mu\nu} =0 \, ,
\ee
since $\eta ={\rm constant}$ is a solution of the equations of motion. Then,  
let the energy in the massive theory be proportional to the space
integral of $\Theta^{\rm broken}_{00}$. Using \rf{smunu} and \rf{teta}, we
obtain the soliton mass in the form 
\br
{M \o{\sqrt{1-v^2}}} &\equiv& -\int_{-\infty}^{\infty} dx \, \Theta^{\rm
broken}_{00} + E_{\rm vac.}  = -{k\, N_{\bf s}\o l} \pa_x \(\nu + \Omega
x_{+}x_{-}\) \mid_{-\infty}^{\infty}\, , 
\lab{solmass}
\er
because the integral of $L_{00}^{\rm red.}$ vanishes by the above arguments. 
Here   
$v$ is the soliton velocity in the units of the speed of the light. Notice 
that we have subtracted the energy $E_{\rm vac.}$ of the vacuum solution which
is, in fact,  divergent. Of course, the vacuum solution is not unique, and it 
is not  clear which one provides the absolute minimum of the energy. We will 
use the following prescription for the soliton mass formula. For the
soliton solutions lying, under the dressing transformations, on the orbit of
the vacuum solution \rf{vacuum2}, we take $\Omega$ in \rf{solmass} to be that
one given in \rf{omega}. However, for those soliton solutions lying on the
orbit of the vacuum \rf{vacuum1}, we take $\Omega$ in \rf{solmass} to be 
equal to the parameter $\b$
introduced in \rf{vaccond}. Such a prescription guarantees the finiteness of 
the soliton masses.

The soliton masses are determined solely by the behaviour at $x=\pm
\infty$ of the space derivative of the field $\nu$. That is quite a remarkable
fact. In addition, as we now explain, it is very easy to obtain
such a behaviour in the general case from the solitonic solutions 
\rf{nicesoliton}.

Consider an integral gradation of $\cgh$, with $s^{\pr}_i = {\psi^2
\o{\a_i^2}}\, s_i$, $\alpha_0\equiv - \psi$, and $s_i$ labeling the gradation 
that
defines the model \rf{eqm1}--\rf{eqm4}.  Consider the integrable highest weight
representation with highest weight state
\be 
\mid\l_{\bf s}^{\pr}\rangle =
\bigotimes_{i=0}^{r} \,  \mid {\hat{\l}}_i\rangle^{\oplus s_i^{\pr}},
\ee
where $\mid {\hat{\l}}_i\rangle$ are the highest weight states of the 
fundamental representations of $\cgh$, and ${\hat{\l}}_i$ are the 
corresponding fundamental weights of $\cgh$. 

Then it is possible to show \ct{SAVGERV} that taking the expectation value of 
both sides of \rf{nice} in
such state, one gets (with the gauge choice $h_L(x_{-})=h_R(x_{+})=1$)
\be
e^{-\( \nu + \Omega x_{+} x_{-}\) N_{\bf s} {\psi^2\o 2}} = \langle \l_{\bf
s^{\pr}} \mid e^{x_{+}\, \ce_{+}} \, e^{-x_{-}\, \ce_{-}} \, \rho \, e^{x_{-}\,
\ce_{-}}\, e^{-x_{+}\, \ce_{+}} \mid \l_{\bf s^{\pr}} \rangle .
\ee
Now, choosing $\rho$ to be the exponential of an eigenvector of $\ce_{\pm}$,
\be
\lb \ce_{\pm} \, , \, V \rb = \omega_{\pm} V,
\lab{eigenvalue}
\ee
we obtain a soliton solution
\be
e^{-\( \nu + \Omega x_{+} x_{-}\) N_{\bf s} {\psi^2\o 2}} = \langle  
\l_{{\bf s}^{\pr}}  \mid e^{e^{\Gamma} \, V} \mid \l_{{\bf s}^{\pr}} \rangle 
\ee
with $\Gamma = \omega_{+} x_{+} - \omega_{-} x_{-} \equiv \gamma \( x - vt\)$.

Suppose $V$ is an operator in such a representation for which  there is
a positive integer $N_V^{\pr}$, such that
\be
\langle \l_{{\bf s}^{\pr}} \mid V^n \mid \l_{{\bf s}^{\pr}} \rangle =0 
\qquad \mbox{\rm for $n>N_V^{\pr}$}.
\lab{truncation}
\ee
Then the soliton mass is easily obtained from \rf{solmass},
where for $\gamma > 0$ 
($\gamma < 0$) only the upper (lower) limit $x=\infty$ ($x=-\infty$)
contributes in the integral\footnotemark{\footnotetext{We point out that the
soliton mass formula \rf{solitonmass} could be equally obtained by defining the
mass  through the momentum formula, instead through the energy like
in \rf{solmass}, as ${M\, v\o{\sqrt{1-v^2}}} \equiv \int \, dx \Theta_{01}^{\rm
broken}$. In this case, we do not have to subtract the vacuum momentum, since
it vanishes.}} ,    
\be
M = {2\o {\psi^2}}{k\, N_V^{\pr}\o l} \mid \gamma \mid {\sqrt{1-v^2}}= 
 {2\o {\psi^2}}\, {2\,k\,\,N_V^{\pr}\o l} \, \sqrt{\omega_{+}\, \omega_{-}}.
\lab{solitonmass}
\ee

Notice that we must have $\omega_{+}\omega_{-}>0$ in order to have the soliton
velocity $v = (\omega_{-}-\omega_{+})/(\omega_{-}+\omega_{+})$,   
not  exceeding the light velocity ($c=1$). 

The soliton mass formula \rf{solitonmass} has some remarkable properties. One
of them concerns the relation  particle--soliton in the theory,
indicating some sort of duality similar to the electromagnetic duality of some
four dimensional gauge theories possessing the Bogomolny (monopole) limit 
\ct{duality}. 
As we have seen, the soliton solutions are created by the eigenvectors $V$ of
$\ce_{\pm}$. From \rf{eigenvalue} one has 
$\lb \ce_{+} \, , \,\lb \ce_{-} \, , \, V \rb\rb  = \omega_{+}\omega_{-} V$. 
Expanding $V$ over the eigenvectors of the grading operator $Q_{\bf s}$ as 
$V = \sum_n V^{(n)}$, one observes that 
$\lb \ce_{+} \, , \,\lb \ce_{-} \, , \, V^{(n)} \rb\rb  = \omega_{+}\omega_{-} 
V^{(n)}$. Therefore, if some $V^{(n)}\in \cgh_n$, $n=0,\pm 1,\pm 2,\ldots \pm
(l-1)$, does not vanish, it implies that $V^{(n)}$ must be one of 
the eigenvectors $X$ in \rf{fundeigen}. Then we associate a soliton with a
fundamental particle. In addition, we have $\lambda
\equiv \omega_{+}\omega_{-}$, and, consequently, from \rf{massf} and
\rf{solitonmass}, the masses of the corresponding soliton and fundamental
particle are determined by  the same eigenvalue. In fact, we have from
\rf{massf} and \rf{solitonmass}, with $v_{\eta}=1$, that
\be
M_{\rm sol.} =  {2\o {\psi^2}}\, {k\,\,N_V^{\pr}\o l}\, m_{\l}^{\rm part.}.
\ee
Of course, in the expansion of $V$, we may have more than one non
vanishing $V^{(n)}$, with $n=0,\pm 1,\pm 2,\ldots \pm (l-1)$. Then we would 
associate a one--soliton solution to more than one fundamental particle. The
counting of one--soliton solutions has to be better analysed in each particular
case. We discuss this issue in section \ref{sec:exsl2}.

\section{The matter fields}
\label{sec:spinors}

It is clear from \rf{ctf1}-\rf{ctf5}, that the
 massive fields associated with non
vanishing grade (namely $F^{\pm}_m$),  are chiral fields with
non vanishing spins, in contrast with the Toda type fields.
In fact, we show that the free equations for such fields take the form of
the massive Dirac equation, as could be expected from
general
covariance arguments.

Consider the subspace $\cgh_m$ for $0<m<l$. Let $\cgh_m^{(F)}$ be the subspace
of $\cgh_m$, generated by the eigenvectors of $\lb E_{-l}\, , \, \lb E_{l}\, , 
\, \cdot \rb \rb$ with non zero eigenvalues, i.e.,
\be
\cgh_m^{(F)} \equiv \{ T^{(m)} \in \cgh_m  \mid  \l^{(m)} \neq 0 \} ,
\ee
where $\l^{(m)}$ is defined as
\be
\lb E_{-l}\, , \, \lb E_{l}\, , \, T^{(m)} \rb \rb = 
\lb E_{l}\, , \, \lb E_{-l}\, , \, T^{(m)} \rb \rb = \l^{(m)} \, T^{(m)}.
\ee
Decompose the subspace $\cgh_m$, as a vector space, into the sum
\be
\cgh_m = \cgh_m^{(F)} + \cgh_m^{(K)},
\ee
where $\cgh_m^{(K)}$ is the complement of $\cgh_m^{(F)}$ in $\cgh_m$.

It is possible to show (see \ct{SAVGERV} for details) that 
the subspaces
$\cgh_{-l+m}^{(F)}$ and $\cgh_{m}^{(F)}$ are isomorphic. The mapping is given
by the action of $E_{-l}$ on $\cgh_{m}^{(F)}$, or equivalently by the action   
of $E_{l}$ on $\cgh_{-l+m}^{(F)}$.
Therefore, we can put in one-to-one correspondence the fields in 
$\cgh_{-l+m}^{(F)}$ and $\cgh_{m}^{(F)}$. Then, one can show that each pair of
such fields constitute a Dirac spinor under the two dimensional Lorentz group,
and their equations of motion can indeed be written in the form of (obviously
not free) Dirac equations. 
Consequently, we interpret the massive fields associated to generators of non
zero grading as matter fields

\sect{An example of a special class of models}
\label{sec:exsl2}

There is a class of models possessing a $U(1)$ Noether current,
which, under some circunstances, is proportional 
 to a topological current. That
 occurs for those models where the grade $l$ of the operator $E_l$, introduced
 in \rf{fp}, is equal to the integer $\ns$ defined in \rf{gradop}. In addition,
it is necessary that the operators $E_{\pm \ns}$ satisfy the condition
\be
z E_{-\ns} = \mu z^{-1} E_{\ns}  \in \mbox{\rm center of $\cgh_0$}
\lab{parallel}
\ee
where, $\mu$ is some constant independent of $z$, and $z$ is a complex variable
 used to realize the generators of the 
 affine Kac-Moody algebra $\cgh$, in terms of those of the finite simple Lie 
algebra $\cg$ as 
\be
H^n_a \equiv z^n \, H_a  \, , \qquad E_{\a}^n \equiv z^n \, E_{\a} 
\, , \qquad D \equiv z {d \,\, \o dz}
\ee
That means that the ``projections'' of $E_{\pm \ns}$ onto $\cgh_0$, are
parallel and lie in the center of $\cgh_0$. 
When those condition are satisfied it is possible to gauge fix the conformal
symmetry, such that a special $U(1)$ Noether charge is proportional to a
topological charge.

In this section we discuss a example where that happens. It correspond to the
principal gradation of  $sl(2)^{(1)}$ with $l=2$. 
Let us denote by $H^n$, $E_{\pm}^n$, $D$ and $C$ the Chevalley basis
generators of the $sl(2)^{(1)}$. The commutation relations are
\br
\lb H^m \, , \, H^n \rb &=& 2 \, m \, C \, \d_{m+n,0}, \qquad 
\lb E^m_{+} \, , \, E^n_{-} \rb = H^{m+n} + m \, C \, \d_{m+n,0},
\lab{sl2a}\\
\lb H^m \, , \, E^n_{\pm} \rb &=& \pm 2 \, E^{m+n}_{\pm}, 
\qquad 
\lb D \, , \, T^m \rb = m \, T^m \, , \quad T^m \equiv H^m , E_{\pm}^m;
\lab{sl2d}
\er
all other commutation relations are trivial.
The grading operator for the principal gradation (${\bf s}=(1,1)$) is
$ Q\equiv \h H^0 + 2 D$. 
Then the eigensubspaces are
$\cgh_0 = \{ H^0, C, Q\}$, 
$\cgh_{2n+1} = \{ E_{+}^n , E_{-}^{n+1}\}$, with $ n\in \IZ$, and  
$\cgh_{2n} = \{ H^n\}$, with $n\in \{ \IZ - 0\}$.

The mapping $B$ is parametrised as
\be
B= e^{\vp\, H^0}\, e^{{\tilde {\nu}} \, C}\,e^{\eta\, Q} =
e^{\vp\, {\tilde H}^0}\, e^{\nu \, C}\,e^{\eta\, Q},
\lab{bsl2}
\ee
where ${\tilde H}^0 = H^0 -\h \, C$ is the Cartan generator in the
special basis introduced in \rf{cartanplus}, and so ${\tilde {\nu}} =
\nu - \h
\vp$.

In the case $l=2$, we choose
\be
E_2 \equiv m \, H^1 \, , \qquad E_{-2} \equiv m \, H^{-1},
\lab{e2}
\ee
where $m$ is a  constant.
We then have
\be
\lb E_{2} \, , \, \lb E_{-2} \, , \, E_{\pm}^n \rb\rb =
4 m^2 \, E_{\pm}^n.
\ee
Therefore, each of the subspaces $\cgh_{\pm 1}$ has two generators with the
same eigenvalue
$ 4 m^2$.
Following section \ref{sec:spinors} we write
\be
F^{+}_1 = 2\sqrt{i m}\( \psi_R\, E_+^0 +
\widetilde \psi_R E_-^1\)\, ,\quad
F^{-}_1 = 2\sqrt{i m}\( \psi_L\, E_+^{-1} -
\widetilde \psi_L\, E_-^0 \) ,
\ee
and introduce the  Dirac fields 
\br
\psi = \(
\begin{array}{c}
\psi_R\\
\psi_L
\end{array}\) \, ; \qquad
\widetilde \psi = \(
\begin{array}{c}
\widetilde \psi_R\\
\widetilde \psi_L
\end{array}\)
\er
{}From \rf{massf} we obtain the masses of the particles,
\be
m_{\vp} = m_{{\tilde{\nu}}} = m_{\eta} = 0;\; \qquad
m_{\psi} = 4 m;
\lab{mass}
\ee
The equations of motion derived   from \rf{eqm1}--\rf{eqm4}, are
\br
\pa^2 \,\vp &=& -4 m_{\psi}\,  \overline
\psi \gamma_5 e^{\eta+2\varphi \gamma_5} \psi,
\lab{sl2eqm1}\\
\pa^2\,{\tilde{\nu}} &=& -  2 m_{\psi}\,  \overline
\psi (1-\gamma_5) e^{\eta+2\varphi \gamma_5} \psi
 - {1\over 2}  m_{\psi}^2 e^{2\eta},
\lab{sl2eqm2}\\
\pa^2 \eta &=& 0,
\lab{sl2eqm3}\\
i \gamma^{\mu} \pa_{\mu} \psi &=& m_{\psi}\, e^{\eta+2\vp\,\gamma_5} \,
\psi ,\lab{sl2eqm4}\\
i \gamma^{\mu} \pa_{\mu} \widetilde \psi &
=& m_{\psi}\, e^{\eta-2\vp\,\gamma_5} \,
\widetilde  \psi ,
 \lab{sl2eqm5}
\er
where the gamma matrices are defined as
\br
\gamma_0 = -i \(
\begin{array}{rr} 0&-1\\ 1&0
\end{array}\) , \qquad 
\gamma_1 = -i \(
\begin{array}{rr}  0&1\\  1&0
\end{array}\),
\lab{gammas}
\er
and $\gamma_5 =\gamma_0\gamma_1$,
and  ${\bar{\psi}} \equiv {\widetilde \psi}^{T} \, \gamma_0$.
Recall that  $\pa^2 = \pa_t^2
- \pa_x^2$, $x_{\pm}=t \pm x$.
The corresponding Lagrangian has the form
\br
{1\o k}\, \cl &=& {1\o 4} \pa_{\mu} \vp \, \pa^{\mu} \vp
+ {1\o 4}  \pa_{\mu} \vp \, \pa^{\mu} \eta
+ \h  \pa_{\mu} {\tilde{\nu}} \, \pa^{\mu} \eta
- {1\o 8}\, m_{\psi}^2 \, e^{2\,\eta} \nonu\\
&+& i  {\bar{\psi}} \gamma^{\mu} \pa_{\mu} \psi
- m_{\psi}\,  {\bar{\psi}} \,
e^{\eta+2\vp\,\gamma_5}\, \psi.
\lab{lagrangian}
\er
It is real (for $\eta = \mbox{\rm real constant}$) if $\widetilde \psi$ is
the complex conjugate of $\psi$, and
if $\varphi $ is pure imaginary. This will be true for the soliton solution
as we shall see below.

Notice that such model is invariant under the  transformations
\be
x_{+} \leftrightarrow x_{-} \, ; \quad \psi_R \leftrightarrow
\epsilon \widetilde \psi_L \, ; \quad  \widetilde \psi_R \leftrightarrow
-\epsilon \psi_L \, ; \quad \vp \leftrightarrow \vp \, ; \quad \eta
\leftrightarrow \eta \, ; \quad \nu \leftrightarrow \nu
\ee
where $\epsilon = \pm 1$. It should be interprated as the product CP
of charge conjugation times parity.  Parity alone is clearly violated.

The generator $H^0 \in \cgh_0$ commutes with $E_{\pm 2}$, and, therefore, the
gauge symmetry \rf{gs1}--\rf{gs3} of the model is $U(1)_L \otimes U(1)_R$,
\be
h_L(x_{-}) = e^{\xi_- (x_{-}) \, H^0}\, , \qquad
h_R(x_{+}) = e^{\xi_+ (x_{+}) \, H^0}.
\lab{gaugesl2}
\ee
Since the genereators of $U(1)_L$ and $U(1)_R$ are the same, we have the
global gauge transfomations \rf{diagonalgauge} generated by $h_D \equiv h_L =
h_R^{-1}\equiv e^{i \theta \, H^0/2}$ ($\theta = {\rm const.}$). The fields
 are transformed as
\be
\psi \ra e^{i\theta} \psi \, \quad
\widetilde \psi \ra e^{-i\theta} \widetilde \psi \, \quad
\vp \ra \vp \, , \quad {\tilde{\nu}}\ra {\tilde{\nu}} \, , \quad
\eta \ra \eta ;
\lab{globalsl2}
\ee
and the corresponding Noether current is
\be
J^{\mu} = {\bar{\psi}}\, \gamma^{\mu}\, \psi \, , \qquad
\pa_{\mu}\, J^{\mu} = 0.
\lab{noethersl2}
\ee
The fields $\psi$ and $\widetilde \psi$ have charges $1$ and $-1$,
respectively; and $\vp$, ${\tilde{\nu}}$ and $\eta$ have charge zero.

Let us next see how the general ar\-gu\-ments gi\-ven abo\-ve con\-cer\-ning
Noe\-ther and topological charges apply here. The topological
current and charges are
\be
j^{\mu} =  {1\o{2\pi i}}\epsilon^{\mu\nu} \pa_{\nu} \, \vp
\, , \qquad
Q_{\rm topol.} \equiv \int \, dx \, j^0 \, , \qquad
\ee
Indeed, the Lagrangian \rf{lagrangian} has infinitely
many degenerate vacua
due to the invariance under $\vp \ra \vp +  i\pi$.
Making use of the field equations, one easily verifies that
\be
\partial_\mu \left[ i\bar \psi \gamma_5 \gamma^\mu \psi+{1\over 2}
\partial^\mu \vp \right ]=0
\lab{PCAC}
\ee
Combining this equation with the conservation of the vector current
$\bar \psi \gamma^\mu \psi$, one deduces that there exist two charges
 defined by
$$
{\cal J}=-i \widetilde \psi_R \psi_R +{1\over 2} \partial_+\vp,
\quad
{\bar {\cal J}}=i \widetilde \psi_L \psi_L +{1\over 2} \partial_-\vp
$$
which satisfy $\partial_-{\cal J}=0$, $\partial_+{\bar {\cal J}}=0$. 
We now make a ``gauging fixing'' of the conformal symmetry by choosing 
${\cal J}={\bar {\cal J}}=0$. We call it  a ``gauging fixing'', because  any   
values of ${\cal J}$ and ${\bar {\cal J}}$ can be
transformed to zero by a conformal transformation. 
This gives, altogether, 
\be
{1\o{2\pi i}}\epsilon^{\mu\nu} \pa_{\nu} \, \vp=
{1\o \pi} \bar \psi \gamma^\mu  \psi,
\lab{currents}
\ee
so that the topological and Noether  currents
are  proportional. As discussed at the beginning of this section, that is
a consequence of the fact that $E_{\pm 2}$ satisfies \rf{parallel}. 

Let us turn to the Noether charge  which here is simply the
fermion number. It should be defined such that
it satisfies the Poisson bracket relation
\be
i \left \{ \psi, Q_{\rm Noether}\right\}_{\rm P.B.}= \psi
\ee
Since the coupling constant $k$ was taken as an overall factor, this is
satisfied by
\be
Q_{\rm Noether}={k} \int dx
\bar \psi \gamma^0  \psi
\ee
so that
\be
Q_{\rm topol.}={1\over k \pi} Q_{\rm Noether}
\lab{top-noe}
\ee
As argued in general, this means that 
$k$ should only take discrete values 
as  expected, since our actions are related
with the one of WZNW.

Let us now construct the soliton solutions. The operators $E_{\pm 2}$ given 
in \rf{e2}, lie in the homogeneous Heisenberg
subalgebra generated by $H^n$, with the commutation relations 
\rf{sl2a}. Such a subalgebra has no generators of grade $\pm 1$ for the
principal gradation. Therefore, the model under consideration has no vacuum
solutions of type \rf{vacuum2}. Then, from \rf{cepm}, we get
\be
\ce_{\pm} = E_{\pm 2} = m \, H^{\pm 1}.
\ee
We perform the dressing transformation starting from the vacuum solution
\rf{vacuum1}, namely
\be
\vp = \eta = \psi=\widetilde \psi  = 0 \, ,  \qquad \, ; \quad
{\tilde{\nu}} = -{1\o 8} m_{\psi}^2 x_{+} x_{-}\equiv \nu_0.
\lab{vacsl2l=2}
\ee
Now, let $\mid\, {\hat{\l}}_0\, \rangle$ and  $\mid\, {\hat{\l}}_1\, \rangle$ 
be
the highest weight states of two fundamental representations of the  affine
Kac--Moody algebra $sl(2)^{(1)}$, respectively the scalar and spinor ones.
Then, from \rf{nice} with $\eta =0$, we obtain the solutions on the orbit of 
the
vacuum  \rf{vacsl2l=2},   
\br
e^{-\vp} &=& {{\langle \,{\hat{\l}}_1\,\mid\, G\, \mid\, {\hat{\l}}_1\,
\rangle}\over  {\langle \,{\hat{\l}}_0\,\mid\, G\, \mid\, {\hat{\l}}_0\,
\rangle}},\; \; \qquad 
e^{-({\tilde{\nu}} - \nu_0)} = \langle \,{\hat{\l}}_0\,\mid\,
G\, \mid\, {\hat{\l}}_0\, \rangle ,\nonu\\
\psi_R &=&  \sqrt{ m\over i}\, {{\langle \,{\hat{\l}}_0\,\mid\, E_-^1\, G\,
\mid\,  {\hat{\l}}_0\, \rangle}\over  {\langle\,{\hat{\l}}_0\,\mid\, G\, \mid\,
{\hat{\l}}_0\, \rangle}},\;\; \;
\widetilde \psi_R =-  \sqrt{m\over i}\, {{\langle
\,{\hat{\l}}_1\,\mid\, E_+^0\, G\, \mid\, {\hat{\l}}_1\, \rangle}\over 
{\langle\,{\hat{\l}}_1\,\mid\, G\, \mid\, {\hat{\l}}_1\, \rangle}}
\nonumber \\ 
\psi_L &=&- \sqrt{m\over i}\, {{\langle \,{\hat{\l}}_1\,\mid\,  G\,E_-^0\,
\mid\, {\hat{\l}}_1\, \rangle}\over  {\langle\,{\hat{\l}}_1\,\mid\, G\, \mid\,
{\hat{\l}}_1\, \rangle}},\; 
\widetilde \psi_L = - \sqrt{m\over i}\,{{\langle \,{\hat{\l}}_0\,\mid\, 
G\,E_+^{-1}\, \mid\, {\hat{\l}}_0\, \rangle}\over 
{\langle\,{\hat{\l}}_0\,\mid\, G\, \mid\, {\hat{\l}}_0\, \rangle}}, 
\lab{solsl2}
\er 
where
\be
G \equiv  e^{x_{+}\, \ce_{+}} \, e^{-x_{-}\, \ce_{-}} \, \rho \, e^{x_{-}\,
\ce_{-}}\, e^{-x_{+}\, \ce_{+}}.
\ee
In order to get the soliton solutions, we choose the fixed mapping
$\rho$ to be an exponentiation of an eigenvector of $\ce_{\pm}$ (solitonic
specialization); namely, $\rho = e^{V}$, with $\lb \ce_{\pm}\, ,\, V \rb = 
\omega_{\pm} \, V$. Therefore,
\be
G = \exp \( e^{\Gamma} \, V\) \quad \mbox{ with } \;
\Gamma = \omega_{+} x_{+} - \omega_{-} x_{-} \equiv \gamma \( x
- v\,t\) .
\ee
In this case the eigenvectors of $\ce_{\pm}$ are
\be
V_{\pm} (z) = \sum_{n \in \IZ} z^{-n}\, E_{\pm}^n.
\ee
Indeed,
\br
\lb \ce_{+}\, ,\, V_{\pm}(z) \rb &=& \pm 2 m z \, V_{\pm}(z) \equiv 
\omega_{+}^{\pm} V_{\pm}(z),\\
\lb \ce_{-}\, ,\, V_{\pm}(z) \rb &=& \pm {2 m\o z }\, V_{\pm}(z) \equiv 
\omega_{-}^{\pm} V_{\pm}(z).
\er
The solution, associated with $V_{+}(z)$, is 
\br
\nu = \nu_0,\, \quad
\vp = \widetilde \psi = 0,\, \quad
\psi = \sqrt{m\over  i}  e^{\Gamma}\, \( 
\begin{array}{r}
z \\
-1
\end{array}\) ;
\lab{sol1}
\er
while those, associated with $V_{-}(z)$, is given by 
\br
\nu = \nu_0,\, \quad
\vp = \psi = 0,\, \quad
\widetilde \psi = -\sqrt{m\over  i} e^{-\Gamma}\, \( 
\begin{array}{r}
1 \\
1/ z
\end{array}\) ,
\lab{sol2}
\er
where
\be
\Gamma = 2 m (z x_+ - {1\o z} x_-)  \equiv \gamma \( x - vt\).
\lab{gammasl2}
\ee
The masses of these solutions are obtained from \rf{solitonmass}. Here the 
relevant 
state $\mid \l_{\bf s^{\pr}}\rangle$ in \rf{truncation} is
\be
\mid \l_{\bf s^{\pr}}\rangle = \mid {\hat{\l}}_{0}\rangle \otimes 
\mid {\hat{\l}}_{1}\rangle .
\lab{lambdaspr}
\ee
Using level one vertex operators, one can verify that
\be
\langle {\hat{\l}}_i \mid \, \( V_{\pm}(z)\)^n \, \mid {\hat{\l}}_i \rangle = 0
\, , \quad \mbox{\rm for $n \geq 1$ and $i=0,1$}.
\ee
Therefore, $N_V^{\pr}=0$ in \rf{truncation}, and from \rf{solitonmass} one
gets that the masses of the solutions \rf{sol1} and \rf{sol2} vanish. 
Such solutions correspond to the objects which travel with velocities 
$v=\pm \( 1 -  z^2\)/\( 1 + z^2\)$; and keeping  
$z^2 >0$, one has $\mid v \mid <1$.
Therefore, these solutions cannot be interpreted as 
solitons (particles), since
they would correspond to massless particles traveling with velocity smaller
that light velocity. We should interpret them as vacuum configurations, since
they have the same energy as vacuum \rf{vacsl2l=2}.

The true soliton solutions of the system are constructed as follows. 
Notice that $V_{+}(z)$ and $V_{-}(-z)$ have the same eigenvalues. Therefore,
any linear combination of them, leads  to solutions traveling with a constant
velocity without dispersion. So, we  let 
\be
V(a_{\pm},z) \equiv \sqrt{i}\left ( a_+ V_{+}(z) + a_- V_{-}(-z)\right);
\lab{nicesol}
\ee
\be
\lb \ce_{+}\, ,\, V(a_{\pm},z) \rb =  2 m z \, V(a_{\pm},z) \, , \qquad 
\lb \ce_{-}\, ,\, V(a_{\pm},z) \rb =  {2 m\o z }\, V(a_{\pm},z),
\ee
and so $\omega_{+}=2 m z$ and $\omega_{-}={2 m\o z }$.
The particular factor $\sqrt{i}$ is chosen such that the reality condition 
will be obeyed with $a_-=a_+^*$. 
Again, using level one vertex operators, one can verify that\footnote{Notice 
that the truncation occurs for powers greater than $4$, and not $2$, because 
$\mid \l_{\bf s^{\pr}}\rangle$ lies in the tensor 
product representation, see 
\rf{lambdaspr}}
\be
\langle \l_{\bf s^{\pr}} \mid \, V(a_{\pm},z)^n \, \mid \l_{\bf s^{\pr}}\rangle
= 0  \qquad \mbox{\rm for $n > 4$}.
\ee
Therefore, $N_V^{\pr}=4$ in \rf{truncation}, and from \rf{solitonmass}
with $\psi^2 =2$, and $\psi$ being the highest root of $sl(2)$,
one gets that the mass of such solutions is 
\be
M= 8 k \, m = 2\, k\, m_{\psi},
\lab{sl2solmass}
\ee
where $k$ is the coupling constant appearing in the Lagrangian \rf{lagrangian}.
The solutions generated by \rf{nicesol}, have two parameters, namely $a_{\pm}$.
One parameter is always present, because one can scale an eigenvector of
$\ce_{\pm}$ without changing the 
width $\gamma$ and velocity $v$ of the soliton,
obtained from the eigenvectors $\omega_{\pm}$; see \rf{nicesoliton}. However,
in this case, the second parameter comes from a symmetry. As we have pointed
out in  \rf{symrho}, associated to the fixed element $\rho =
e^{V(a_{\pm},z)}$, we have an orbit of equivalent solutions due to the global
transformations \rf{globalsl2},
\be
V(a_{\pm},z) \ra \sqrt{i}\left ( 
a_+\, e^{i\theta}\, V_{+}(z) + a_-\, e^{-i\theta}\, V_{-}(-z)\right) .
\ee
The explicit form of the solutions generated by \rf{nicesol}, is obtained 
using \rf{solsl2}, 
\br
\vp &=& \log \(  {{1 + i\sigma e^{2 \Gamma}}\o { 1 - i\sigma e^{2 \Gamma}}}\) ,
\lab{solsl2a}\\
{\tilde {\nu}} &=& - \log \( 1 + i\sigma e^{2 \Gamma}\) - {1\o 8} m_{\psi}^2 
 x_{+} x_{-},
\lab{solsl2b}\\
\eta &=& 0;
\lab{solsl2c}
\er
and
\br
\psi = a_{+}\sqrt{m} \, e^{\Gamma}\, \(
\begin{array}{c}
 {z\o{1 + i\sigma e^{2 \Gamma}}}\\
{-1 \o{1 - i\sigma e^{2 \Gamma}}}
\end{array}\) \, , \qquad 
\widetilde \psi = a_{-}\sqrt{m}\, e^{\Gamma}\, \(
\begin{array}{c}
 {z\o{1 - i\sigma e^{2 \Gamma}}}\\
{-1 \o{1 + i\sigma e^{2 \Gamma}}}
\end{array}\) ;
\lab{solsl2d}
\er
where $\Gamma$ is given in \rf{gammasl2}, and $\sigma = a_{+} a_{-}z/4$.
Keeping $m$ and $z$ real,  we have the
mass $M$  of the soliton, from \rf{sl2solmass}, 
real and positive, and also the 
parameters $\gamma$ and $v$ \rf{gammasl2} are real. The reality condition is 
obeyed if $a_-=a_+^*$, as anticipated.  At this point, it is useful 
to re-express the expressions just given in terms of the 
physical parameters of the 
soliton. Using equations \rf{gammas} and \rf{mass} and one deduces that 
\be 
\gamma= m_\psi\left / \sqrt{1-v^2}\right., \quad   
z=  \sqrt{ (1-v)/(1+v)}. 
\lab{prm}
\ee
Moreover, since $a_\pm$ are complex conjugate, we may write 
\be
a_\pm =e^{\pm i\theta} 2 \sqrt {\sigma\over z}. 
\lab{apm}
\ee
   The dependence upon space-time 
 appears   only 
through  $\sqrt {\sigma} \exp(\Gamma)$. We will  
write \footnote{by convention, we choose $\sigma$ to be positive}
\be
\sqrt{\sigma} e^{\Gamma}=\exp( (\gamma(x-x_0-vt))
\ee
where $x_0$ is the position of the soliton at 
time zero. 
 Then  we have
\be
\vp = 2i \arctan \( \exp \( 2m_\psi \( x-x_0-vt\)/\sqrt{1-v^2}\)\) ,
\lab{solfinale}
\ee
which is the sine--Gordon soliton. The Dirac fields are given by  
\be
\psi = e^{i\theta} \sqrt{m_\psi} \, 
e^{m_\psi \( x-x_0-vt\)/\sqrt{1-v^2}}\, \(
\begin{array}{c}
\left( { 1-v\o 1+v}\right)^{1/4}  
{1 \o 1 + ie^{2 m_\psi \( x-x_0-vt\)/\sqrt{1-v^2}}}\\
-\left( { 1+v\o 1-v}\right)^{1/4}  
{1 \o 1 - ie^{2 m_\psi \( x-x_0-vt\)/\sqrt{1-v^2}}}, 
\end{array}\) 
\lab{solsimple}
\ee
and $\widetilde \psi$ is the complex conjugate of $\psi$. Thus the only 
parameters are the soliton mass and velocity, 
together with the angle $\theta$ which reflects the global invariance 
\rf{globalsl2}. 
Notice that the sign of the tolopogical charge
 can be reversed by reversing the sign of
$z$. Therefore, the solutions \rf{solfinale}--\rf{solsimple} contain the
sine--Gordon soliton and anti--soliton.

Finally, we come to the  very important feature of the present model already 
mentioned above in general, namely 
it is clear from the explicit expressions Eqs.\ref{solsimple} that $\psi$ 
vanishes 
exponentially when $x-x_0\to \pm  \infty$, so that the Dirac field is 
confined inside the soliton. That this must be true is of course a general 
consequence of Eq.\ref{currents} which may be verified directly on 
the explicit solution. This phenomenon has been much studied for 
electron phonon systems. Models of a similar type  describe  the  
 electron self-localization 
in quasi-one-dimensional dielectrics (for recent reviews see 
\ct{BK}, \ct{HKSW}). At low temperature these systems 
go over to   dielectric 
states  characterized by charge density waves which can 
be constructed on the basis of the Peierls model. The  continuous limits  
are  described by Lagrangians 
similar to Eq.\ref{lagrangian}. Discussing this important issue  
is beyond the scope of the present article, 
so we will not dwell upon it here. 
Let us simply recall that the typical example of the polyacteline molecule 
was much discussed in connection with fermion number 
fractionization \ct{JR}.  Clearly, on the other hand one may regard 
our soliton solution  a sort of one dimensional bag model for QCD. 
In this connection let us note that, if we introduce the two-by-two 
matrix $U=\exp(\eta+2\varphi \gamma_5)$, we may rewrite the 
Lagrangian Eq.\ref{lagrangian} as 
\br
\cl &=& {1\over 16} \Bigl \{ \hbox{tr} \left [ U^{-1} \partial_\mu U
 {1+\gamma_5\over 2}  U^{-1} \partial^\mu U\right]
-{1\over 2} \hbox{tr} \left [ U^{-1} \partial_\mu U
 \right]\hbox{tr} \left [  U^{-1} \partial^\mu U\right]\Bigr \}
\nonu\\
& &
+i\bar \psi \gamma_\mu \partial_\mu \psi
- \bar \psi  U \psi 
-{m_\psi^2\over 8} \hbox{det} (U), 
\lab{langrU}
\er
which is similar to a two-dimensional version of the low energy effective 
action for QCD (see e.g. \ct{CHT}).

\sect{Conclusions} 

From the common description based on  zero curvature equations, a unified 
treatment has been given of the procedures to obtain solutions of  nonlinear 
equations for a large class of integrable hierarchies including the 
 generalized KdV and non-abelian Toda coupled to matter. 
 Interesting examples of them have been discussed in detail, taking 
advantage of the simplifications achieved, and the same procedure can be
equally well applied to more complex 
equations. The role of  the tau functions has been clarified, as specific 
matrix elements in special integrable highest weight representations 
(for any level). The 
results are a further step towards stablishing the conjecture that all 
multisoliton solutions lie in the orbit of the vacuum, generated by dressing 
transformations, which are now clearly related to the tau functions and 
solitonic specialization.
These results should be very useful for the classification of integrable 
theories
in two dimensions and for the generalization to the quantum case and to higher 
dimensions.   

\vspace{.5 cm}

\noindent{\bf Acknowledgements}

The authors are very grateful to the organizers of SIMI/96 for the invitation,
and for the very kind hospitality.

\end{document}